\documentclass[preprint,12pt,authoryear]{elsarticle}
\usepackage{amsmath}
\usepackage{amssymb}
\usepackage{booktabs}
\usepackage{lineno,hyperref}
\usepackage{float}
\modulolinenumbers[1]

\newcommand{\RN}[1]{%
  \textup{\uppercase\expandafter{\romannumeral#1}}%
}

\newtheorem{theorem}{Theorem}
\usepackage{graphicx}
\usepackage[caption=false]{subfig}
\usepackage{lineno,hyperref}
\modulolinenumbers[5]

\usepackage{graphicx}
\usepackage{amssymb}
\usepackage{lineno}
\journal{}

\begin{document}

\begin{frontmatter}


\title{Characteristic, completion or matching timescales? An analysis of temporary boundaries 
in enzyme kinetics}

\author[label1]{Justin Eilertsen}

\author[label1]{Wylie Stroberg}

\author[label1,label2]{Santiago Schnell\fnref{cor1}}
\ead{schnells@umich.edu}

\address[label1]{Department of Molecular \& Integrative Physiology, University of Michigan 
Medical School, Ann Arbor, MI 48109, USA}
\address[label2]{Department of Computational Medicine \& Bioinformatics, University of 
Michigan Medical School, Ann Arbor, MI 48109, USA}

\fntext[cor1]{Corresponding author}

\begin{abstract}
Scaling analysis exploiting timescale separation has been one of the most important 
techniques in the quantitative analysis of nonlinear dynamical systems in mathematical
and theoretical biology. In the case of enzyme catalyzed reactions, it is often overlooked 
that the characteristic timescales used for the scaling the rate equations are not ideal 
for determining when concentrations and reaction rates reach their maximum values. In this 
work, we first illustrate this point by considering the classic example of the single-enzyme, 
single-substrate Michaelis--Menten reaction mechanism. We then extend this analysis to a more 
complicated reaction mechanism, the auxiliary enzyme reaction, in which a substrate is 
converted to product in two sequential enzyme-catalyzed reactions. In this case, depending 
on the ordering of the relevant timescales, several dynamic regimes can emerge. In addition 
to the characteristic timescales for these regimes, we derive matching timescales that 
determine (approximately) when the transitions from initial fast transient to steady-state 
kinetics occurs. The approach presented here is applicable to a wide range of singular 
perturbation problems in nonlinear dynamical systems.
\end{abstract}


\end{frontmatter}

\linenumbers

\section{Introduction} \label{S:1}
Nonlinear differential equations are used to model the dynamical behavior of natural 
phenomena in science. As the natural phenomena become more complex, the dynamics are 
influenced by multiple timescales, which create technical problems in the mathematical 
analysis and numerical computation of models \citep{Lin:1988:MAD}.  

The 21st century has been dominated by advances in the biological and biomedical sciences. 
As a result, examples of complex dynamical systems have become ubiquitous in theoretical 
and mathematical biology. Despite their complexity, all major levels of biological organization 
have one common dynamical denominator: chemical reactions are continuously taking place in 
living systems. Most of these reactions involve enzymes. Arguably, if biology is to be 
understood as a dynamical process, enzyme catalyzed reactions need to be investigated 
quantitatively \citep{Gallagher:2004:EMW}.

The quantitative description of any enzyme catalyzed chemical reaction is often decomposed 
into two categories: thermodynamics and kinetics. The former tells us if a particular reaction 
is favorable, while latter describes the timescales over which reactions occur. From the point 
of view of the experimental scientist, chemical kinetics focuses on the measurement of 
concentrations as a function of time with the goal of characterizing reaction 
properties \citep{Espenson:1995:CKR}. Regardless of whether a kinetic model is linear or 
nonlinear, stochastic or deterministic, the effectiveness of the model is only as good as the 
timescales it predicts \citep{Shoffner:2017:AET}: timescales provide not only an estimation 
of the effective duration of the reaction, but are also critical in characterizing reaction
mechanisms. This topic is not unfamiliar to Philip K. Maini, who has worked in a number of 
diverse areas of mathematical biology, including enzyme 
kinetics \citep{Frenzen1988,BURKE199081,Burke199397,SCHNELL2000483,SCHNELL2002137,article}.

Philip K. Maini mentored one of us, Santiago Schnell, through the rigorous theory of timescale 
analysis in chemical kinetics that lies at the intersection of chemistry and geometric singular
perturbation theory (GSPT). In fact, GSPT is widely applicable not only to chemical kinetics, 
but to a plethora of important biological models \citep{Bertram2017}. Largely, GSPT is the study 
of dynamical systems of the form
\begin{subequations}\label{eq:SF}
\begin{align}
\dot{x}&=f(x,y),\\
\varepsilon\dot{y}&=g(x,y),
\end{align}
\end{subequations}
where $\dot{\phantom{x}}$ and $\varepsilon \ll 1$ denote differentiation with respect to time.
These systems are often referred to as slow/fast systems, since changes in the variable $x$ occur 
over timescales that are large compared to the timescales over which the variable $y$ changes. 
For example, if time is rescaled as $t_{\varepsilon} = t/\varepsilon$, then the evolution of 
(\ref{eq:SF}) becomes
\begin{subequations}\label{eq:SF1}
\begin{align}
x'&= \varepsilon f(x,y),\\
y'&= g(x,y),
\end{align}
\end{subequations}
with $'$ denoting differentiation with respect to $t_{\varepsilon}$. Over the $t_{\varepsilon}$-timescale, 
the variable $x$ barely changes, while the variable $y$ can change significantly. In contrast, 
the change in variable $x$ is nontrivial over the $t$-timescale and, due to the presence of 
slow manifolds \citep{Roussel:1990:GSSA}, the change in the variable $y$ can be shown to be explicitly 
dependent on change in $x$. Thus, the dynamics of (\ref{eq:SF}) is dependent on two different timescales: 
the fast timescale, $t_{\varepsilon}$, and the slow timescale, $t$. Each timescale defines a unique 
dynamical regime: the initial, ``$t_{\varepsilon}$-regime'', over which $x$ is essentially constant 
and $y$ changes rapidly, and the ``$t$-regime'', in which $x$ changes significantly and the 
change in $y$ is dependent on the change in $x$. 

GSPT has a rich relationship with chemical kinetics, particularly regarding the application of 
\textit{matched asymptotics}. Matched asymptotics is a common mathematical approach aimed at 
finding an accurate approximation to the solution of an equation, or system of 
equations \citep[see][for an excellent discussion on matched asymptotics]{kuehn2015multiple}. 
Usually, the study of matched asymptotics is linked to singular perturbation problems that arise 
as a consequence of underlying disparate spatial layers, such as boundary layers that form in 
pattern formation during embryonic development \citep[see][]{Maini:2012:TMB}. The 
specific aim of matched asymptotics is to generate a \textit{composite solution}, which is 
constructed by gluing together local solutions (solutions that are asymptotically valid on 
different regimes) to comprise a solution that is uniformly valid \citep{Holmes:2013:IPM}. Of 
principal interest in chemical kinetics, for which there typically exist multiple disparate 
timescales, is to determine the timescales that contribute to the composite solution. 

In this work, we begin by introducing the characteristic timescale, which is a well-defined
timescale from dynamical systems theory. We show that the established ``fast timescale" of 
the single-enzyme, single-substrate, Michaelis--Menten (MM) reaction mechanism is in fact a
characteristic timescale, and we demonstrate that characteristic timescales do provide a 
correct ``partitioning" of time into the different slow and fast sub-domains from which the 
composite solution should be constructed. However, we also show that characteristic
timescales are not suitable for determining \textit{when} a transition from one 
dynamical regime to another dynamical regime occurs. This means that characteristic timescales
cannot tell us when concentrations of certain chemical species reach their peak values, 
or when the rate of product generation reaches its maximum value. Thus, there is 
a need for an additional timescale, which we refer to as a \textit{matching timescale}, 
that provides a temporal boundary between specific dynamic (kinetic) regimes. Its derivation 
follows directly from the theory of GSPT and matched asymptotics, and we demonstrate that 
appropriate matching timescales can be constructed from physical knowledge of the characteristic 
timescales. Specifically, through the application of Tikhonov--Fenichel 
Theory~\citep{tikhonov1952,Fenichel1971}, we derive the correct matching timescale for the 
MM reaction mechanism, and show that it can be explicitly obtained from the fast and slow 
characteristic timescales. We also categorize the corresponding slow timescale of the MM 
reaction mechanism as either a characteristic, depletion, or completion timescale. 

Most chemical reactions that consist of two disparate timescales are well-understood. However, 
much of the modern GSPT analyzes problems comprising \textit{more} than two 
timescales~\citep{BVW,Rubin,LRV}, and it is time to push enzyme kinetics in this direction. 
Thus, in this work, we analyze the kinetics of the auxiliary enzyme reaction 
mechanism \citep{Eilertsen:2018:KAS}
\begin{align}
\nonumber S_1 + E_1&   \begin{array}{c} {\scriptstyle k_1}\\
                       \rightleftharpoons\\
                       {\scriptstyle k_{-1}}
        \end{array} C_1 
        \begin{array}{c}
        {\scriptstyle k_2}\\
        \rightarrow \\
        {}
        \end{array} E_1 + S_2, \ 
\end{align}
\begin{align} 
\nonumber S_2 + E_2&   \begin{array}{c} {\scriptstyle k_3}\\
                       \rightleftharpoons\\
                       {\scriptstyle k_{-3}}
        \end{array} C_2 
        \begin{array}{c}
        {\scriptstyle k_4}\\
        \rightarrow \\
        {}
        \end{array} E_2 + P, \ 
\end{align}
under the assumption that the auxiliary enzyme, $E_2$, is in excess with respect to $E_1$. We 
show that there are four timescales in a certain parameter regime of this reaction, and we 
illustrate that different orderings of the timescales must be considered in order to establish 
a complete description of the kinetics. The relevant characteristic timescales that approximate 
the duration of each regime are derived through geometric analysis of the phase-plane. Lastly,
composite solutions and matching timescales are obtained.

\section{The characteristic timescale}
Consider a general, autonomous dynamical system of the form
\begin{equation}\label{eq:gdyn}
\dot{x} = f(x),
\end{equation}
and suppose $f(x)$ has a fixed point, $x^{\ast}$, such that $f(x^{\ast})=0$. The 
\textit{characteristic timescale} is the reciprocal of the exponential growth/decay rate of the 
linearized equation in a small neighborhood surrounding $x^{\ast}$. That is, if $\delta$ is 
a small perturbation, then
\begin{equation}
f(x^{\ast} + \delta) \simeq \cfrac{df}{dx}\bigg|_{x=x^{\ast}} \equiv f'(x^{\ast}),
\end{equation}
and therefore
\begin{equation}
\dot{\delta} \simeq f'(x^{\ast})\delta.
\end{equation}
Since linearized evolution of the perturbation grows or decays according to 
\begin{equation}
\delta \simeq \displaystyle \exp\left[{\displaystyle f'(x^{\ast})t}\right],
\end{equation}
the characteristic timescale, $t_{\chi}$, is the time required for the perturbation to 
\textit{significantly} grow or decay:
\begin{equation}
t_{\chi} = \cfrac{1}{|f'(x^{\ast})|}.
\end{equation}
For a linear, exponential decay differential equation of the form
\begin{equation}
\dot{x} = -\gamma x,\quad x(0)=x_0,
\end{equation}
the characteristic timescale is $1/\gamma$, and corresponds to the exact amount of time it 
takes the initial condition to decay to 
\begin{equation}
x(t_{\chi}) = (1-\ell)x_0, \quad \ell = \cfrac{\exp(1)-1}{\exp(1)},
\end{equation}
which is roughly $0.37x_0$. In addition, for a linear equation of the form
\begin{equation}
\dot{x} = -\gamma x + A, \quad x(0) = 0,
\end{equation}
where $A$ is a constant, the characteristic timescale, $1/\gamma$, is the exact amount of 
time it takes $x$ to grow to
\begin{equation}
x(t_{\chi}) = \ell \cfrac{A}{\gamma} \equiv \ell x^{\max},
\end{equation}
or roughly $0.63 x^{\max}$.

\section{The slow and fast timescales of the Michaelis--Menten reaction mechanism: An exercise 
in the power and limitations of characteristic timescales}
We continue by reviewing the pertinent characteristic timescales for the well-studied 
single-enzyme, single-substrate reaction mechanism (\ref{eq:react1}), in which an enzyme, 
$E_1$, binds to a substrate, $S_1$ (forming an intermediate enzyme-substrate complex, $C_1$), 
and catalyzes the conversion of $S_1$ into product, $P$:
\begin{align} \label{eq:react1}
S_1 + E_1&   \begin{array}{c} {\scriptstyle k_1}\\
                       \rightleftharpoons\\
                       {\scriptstyle k_{-1}}
        \end{array} C_1 
        \begin{array}{c}
        {\scriptstyle k_2}\\
        \rightarrow \\
        {}
        \end{array} E_1 + P. \ 
\end{align}
The kinetics of the reaction depend not only in the rate constants, $k_1$ and $k_{-1}$, and 
the catalytic constant $k_2$, but also on the initial concentrations of $S_1$ and $E_1$. 
Specifically, the \textit{reduced} mass action equations that govern the kinetics 
of (\ref{eq:react1}) are
\begin{subequations}\label{eq:2}
\begin{align}
\dot{s}_1 &= -k_1(e_1^0-c_1)s_1+k_{-1}c_1,\label{eq:MA01}\\
\dot{c}_1 &= k_1(e_1^0-c_1)s_1-(k_{-1}+k_2)c_1.\label{eq:MA02}
\end{align}
\end{subequations}
In this system, $s_1$ denotes the concentration of $S_1$, $c_1$ denotes the concentration 
of $C_1$, and $s_1^0$ and $e_1^0$ are, respectively, the initial substrate and enzyme 
concentrations. The mass action equations~(\ref{eq:MA01})--(\ref{eq:MA02}) can be approximated 
with the differential-algebraic equation,
\begin{subequations}\label{eq:2A}
\begin{align}
\dot{s}_1 &= -\cfrac{k_2k_1}{k_{-1}+k_2+k_1 s_1}s_1,\label{eq:MA0A}\\
c_1 &= \cfrac{k_1e_1^0}{k_{-1}+k_2+k_1s_1}s_1,\label{eq:MA0B}
\end{align}
\end{subequations}
by assuming the quasi-steady-state approximation (QSSA).

Despite a significant body of literature dedicated to the development of methods and 
techniques for estimating timescales in chemical kinetics \citep{Rice,PALSSON1984273,PALSSON1985231,PALSSON1987447,Segel1988,Segel1989,Shoffner:2017:AET},
timescale estimation remains \textit{ad hoc} in most applications, and we will later 
see that this work is no exception. We will study and review (\ref{eq:react1}) in regimes 
where the QSSA is valid. Historically, the most common method employed to study the validity 
of the QSSA is scaling combined with singular perturbation analysis. Early 
studies \citep{HEINEKEN196795} of the validity of the QSSA suggested that the initial 
enzyme concentration must be small in comparison to the initial substrate concentration: $\bar{\varepsilon} \equiv e_1^0/s_1^0 \ll 1$. One of the first authors to recognize that $\bar{\varepsilon} \ll 1$ was an incomplete condition for the validity of the QSSA was
Bernhard Palsson \citep{PALSSON1984273,PALSSON1987447}. Palsson made two important 
discoveries: (1) he recognized that the QSSA was still applicable when $e_1^0 \approx s_1^0$ 
as long as $e_1^0 \ll (k_{-1}+k_2)/k_1$; (2) he noted that the QSSA is \textit{still} 
valid when $e_1^0 \approx s_1^0 \approx (k_{-1}+k_2)/k_1$ as long as 
$\kappa_1 \equiv k_{-1}/k_2 \gg 1$. About a year later, \citet{Segel1988}, who understood 
that there was subtle difference between non-dimensionalization and scaling, correctly 
estimated the disparate timescales of complex formation and substrate depletion. In 
short, the earlier studies failed to determine necessary conditions for the validity 
of the QSSA because, although time had been properly non-dimensionalized in previous 
analyses, it had not been appropriately \textit{scaled}. Thus, history tells us that 
it is difficult, if not impossible, to determine necessary conditions for the validity 
of reduction techniques (like the QSSA) when slow and fast timescales are unknown. 
We will review Segel's analysis in the forthcoming subsections. In addition, we will 
show that the timescales derived by Segel can be used to approximate the matching 
timescale, which gives a better estimation of the time it takes for the reaction to 
reach quasi-steady-state (QSS). 

\subsection{The characteristic initial fast transient of the reaction}
\begin{figure}[htb!]
\centering
\includegraphics[width=12cm]{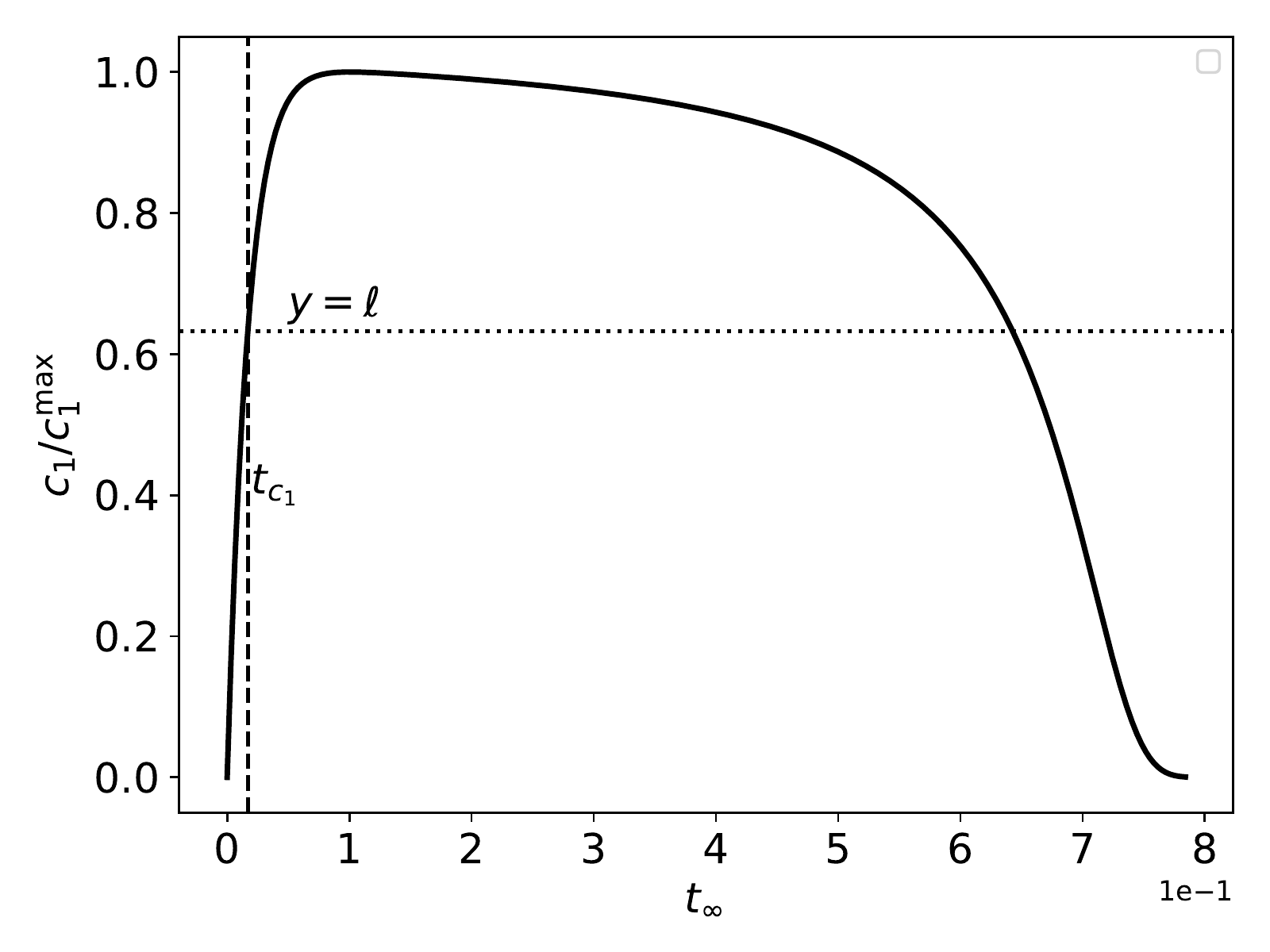}
\caption{\textbf{The validity of $t_{c_1}$ for the Michaelis-Menten reaction 
mechanism~(\ref{eq:react1})}. The solid black curve is the numerically-computed solution 
to (\ref{eq:MA01})--(\ref{eq:MA02}). The dashed vertical curve is corresponds to 
$t_{c_1}= [k_1(K_{M_1}+s_1^0)]^{-1}$. The dotted horizontal line corresponds to 
$\ell c_1/c_1^{\max}=\ell$. The initial concentrations and rate constants used in 
the numerical simulation are: $k_1=0.1$, $k_2=10$, $k_{-1}=1$, $e_1^0=1$ and 
$s_1^0=100$ (units have been omitted). Time has been mapped to the $t_{\infty}$ 
scale: $t_{\infty}(t) = 1-1/\ln[t+\exp(1)]$.} 
\label{fig:1}
\end{figure}

It is well-established that, under the reactant stationary 
assumption~\citep[RSA,][]{Hanson:2008:RSA,Schnell2014}, the dynamics of (\ref{eq:react1}) 
initialize with a brief initial transient during which the intermediate 
complex concentration, $c_1$, accumulates rapidly towards its maximum while the 
substrate $s_1$ remains effectively unchanged from the initial substrate concentration 
$s_1^0$. The RSA ensures $s_1 \approx s_1^0$ during the initial transient of the 
reaction. Under the RSA, equation~(\ref{eq:MA02}) is approximately
\begin{equation}
\dot{c}_1 \simeq k_1(e_1^0-c_1)s_1^0 - (k_{-1}+k_2)c_1,
\end{equation}
which admits the solution
\begin{equation}\label{eq:linear}
c_1 \simeq c_1^{\max}\left[1-\exp(-k_1(K_{M_1}+s_1^0)t)\right], 
					\quad c_1^{\max}=\cfrac{e_1^0s_1^0}{K_{M_1}+s_1^0}.
\end{equation}
In the above equation, $K_{M_1}=(k_{-1}+k_2)/k_1$ is the Michaelis constant. The 
characteristic timescale of the intermediate complex that arises from (\ref{eq:linear}) 
is $t_{c_1}$:
\begin{equation}
t_{c_1}= \cfrac{1}{k_1(K_{M_1}+s_1^0)}.
\end{equation}
Technical justification for $t_{c_1}$ was originally obtained by \citet{Segel1988} and
\citet{Segel1989}. Through scaling analysis, they introduced the dimensionless parameters
\begin{equation}
\sigma_1 \equiv \cfrac{s_1^0}{K_{M_1}},\quad \kappa_1 \equiv k_{-1}/k_2, \quad \beta_1\equiv \cfrac{1}{1+\sigma_1} <1,\quad \alpha_1 \equiv \cfrac{\kappa_1}{1+\kappa_1} < 1,\quad
\end{equation}
allowing equations~(\ref{eq:MA01})--(\ref{eq:MA02}) to be rescaled into their dimensionless form
\begin{equation}\label{eq:5}
\begin{aligned}
\cfrac{d\hat{s}_1}{d\tau}&=\varepsilon_1\left [-\hat{s}_1 + (1-\beta_1)\hat{c}_1\hat{s}_1 + \beta_1\alpha_1\hat{c}_1\right],\quad \varepsilon_1=\cfrac{e_1^0}{K_{M_1}+s_1^0}\\
\cfrac{d\hat{c}_1}{d\tau}&= \hat{s}_1 - (1-\beta_1)\hat{c}_1\hat{s}_1 - \beta_1\hat{c}_1,\\
\end{aligned}
\end{equation}
where $\tau = t/t_{c_1}, \hat{s}_1 = s_1/s_1^0$ and $\hat{c}_1 = c_1/c_1^{\max}$. It 
is clear from (\ref{eq:5}) that if $\varepsilon_1 \ll 1$, then $s_1 \simeq s_1^0$ when 
$t \leq t_{c_1}$. Formally, the qualifier $\varepsilon_1 \ll 1$ \textit{is} the condition 
for the RSA, and $t_{c_1}$ is the \textit{characteristic} timescale of the initial fast 
transient (see {\sc Figure}~\ref{fig:1}).

\subsection{The slow timescale of the MM reaction: from characteristic to completion}
In contrast to the brief timescale over which $c_1$ accumulates (i.e, $t_{c_1}$), 
$s_1$ changes over a much longer timescale. The timescale over which there is appreciable 
change in $s_1$ is the slow timescale of the reaction or the substrate depletion timescale. 
As a direct result of singular perturbation theory, the depletion of $s_1$ is approximately
\begin{equation}\label{eq:redS1}
\dot{s}_1 \simeq -\cfrac{V_1}{K_{M_1}+s_1}s_1
\end{equation}
\textit{after} the initial fast transient (i.e. for $t > t_{c_1}$). The above expression, 
obtained from the QSSA, is known as the MM equation \citep[see,][for reviews]{article,Schnell2014}, 
and $V_1=k_2 e_1^0$ is the limiting rate of the reaction. The slow timescale, $t_{s_1}$, 
is given by
\begin{equation}\label{eq:slowT}
t_{s_1} = \cfrac{s_1^0}{\max |\dot{s}_1|} = \cfrac{K_{M_1}+s_1^0}{V_1}.
\end{equation}
The technical justification of (\ref{eq:slowT}) is acquired through scaling analysis. By 
writing the dimensionless form (\ref{eq:MA01})--(\ref{eq:MA02}) with respect to 
$T = t/t_{s_1}$ yields
\begin{equation}\label{eq:QSSA}
\begin{aligned}
\cfrac{d\hat{s}_1}{dT}&=(1+\kappa_1)(1+\sigma_1)\left [-\hat{s}_1 + (1-\beta_1)\hat{c}_1\hat{s}_1 + \beta_1\alpha_1\hat{c}_1\right],\\
\varepsilon_2 \cfrac{d\hat{c}_1}{dT}&= \hat{s}_1 - (1-\beta_1)\hat{c}_1\hat{s}_1 - \beta_1\hat{c}_1. 
\end{aligned}
\end{equation}
The dimensionless parameter, $\varepsilon_2$, is the ratio of fast and slow timescales: 
$\varepsilon_2 = t_{c_1}/t_{s_1}$.

\begin{figure}[hbt!]
\centering
\includegraphics[width=12cm]{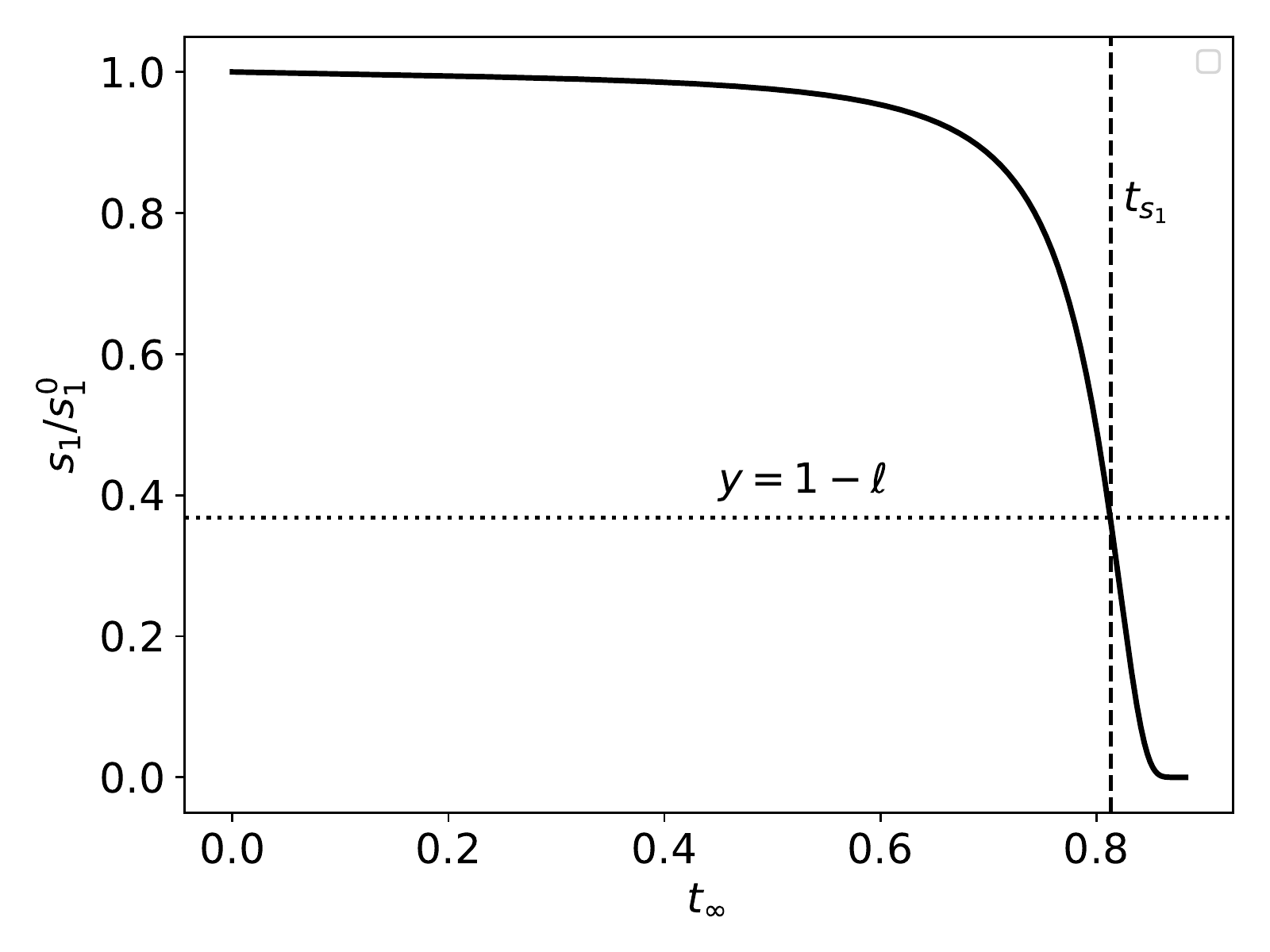}
\caption{\textbf{The graphical illustration of the characteristic timescale for the 
Michaelis--Menten reaction mechanism (\ref{eq:react1}).} When $\sigma_1 \ll 1$, the 
timescale $t_{s_1}$ is the characteristic time of the substrate species. The solid 
black curve is the numerical solution to the mass action 
equations~(\ref{eq:MA01})--(\ref{eq:MA02}) and the vertical dashed/dotted line corresponds 
to $t=t_{s_1}$. The dotted horizontal line corresponds to the scaled characteristic value
$(1-\ell)s_1^0$. The constants (without units) used in the numerical simulation are: $e_1^0=1,k_1=0.01,k_2=10, k_{-1}=1$ and $s_1^0=100$. Time has been mapped to the 
$t_{\infty}$ scale: $t_{\infty}(t) = 1-1/\ln[t+\exp(1)]$.}
\label{fig:2}
\end{figure}

While mathematicians typically refer to $t_{s_1}$ as the slow timescale, the chemical 
interpretation of $t_{s_1}$ depends on the initial \textit{specific concentration}, 
$\sigma_1$. First, the MM equation~(\ref{eq:redS1}) admits a closed-form solution 
with $s_1(t=0)=s_1^0$
\begin{equation}\label{eq:LAM}
s_1 = K_{M_1}W\left[\sigma_1 \exp(\sigma_1 -\eta_1 t)\right] , \quad \eta_1 = \cfrac{V_1}{K_{M_1}},
\end{equation}
where $W\left[\cdot\right]$ is the Lambert-$W$ function \citep{Corless1996,Schnell1997}, 
and the closed-form solution is known as the Schnell--Mendoza 
equation~\citep{Clark:2011:EGU,Feng:2014:RRC,Son:2015:MIP,Murugan:2018:TRE}. If 
$\sigma_1 \ll 1$, then (\ref{eq:LAM}) is asymptotic to
\begin{equation}
s_1 \simeq s_1^0 \exp(\sigma_1 -\eta_1t),
\end{equation}
from which we obtain:
\begin{equation}
s_1(t_{s_1}) \simeq (1-\ell)s_1^0.
\end{equation}
Thus, if the initial substrate concentration is much less than the Michaelis constant, 
$K_{M_1}$, then the slow timescale, $t_{s_1}$, \textit{is} a characteristic timescale 
for the substrate species (see {\sc Figure}~\ref{fig:2}).

\begin{figure}[hbt!]
\centering
\includegraphics[width=12cm]{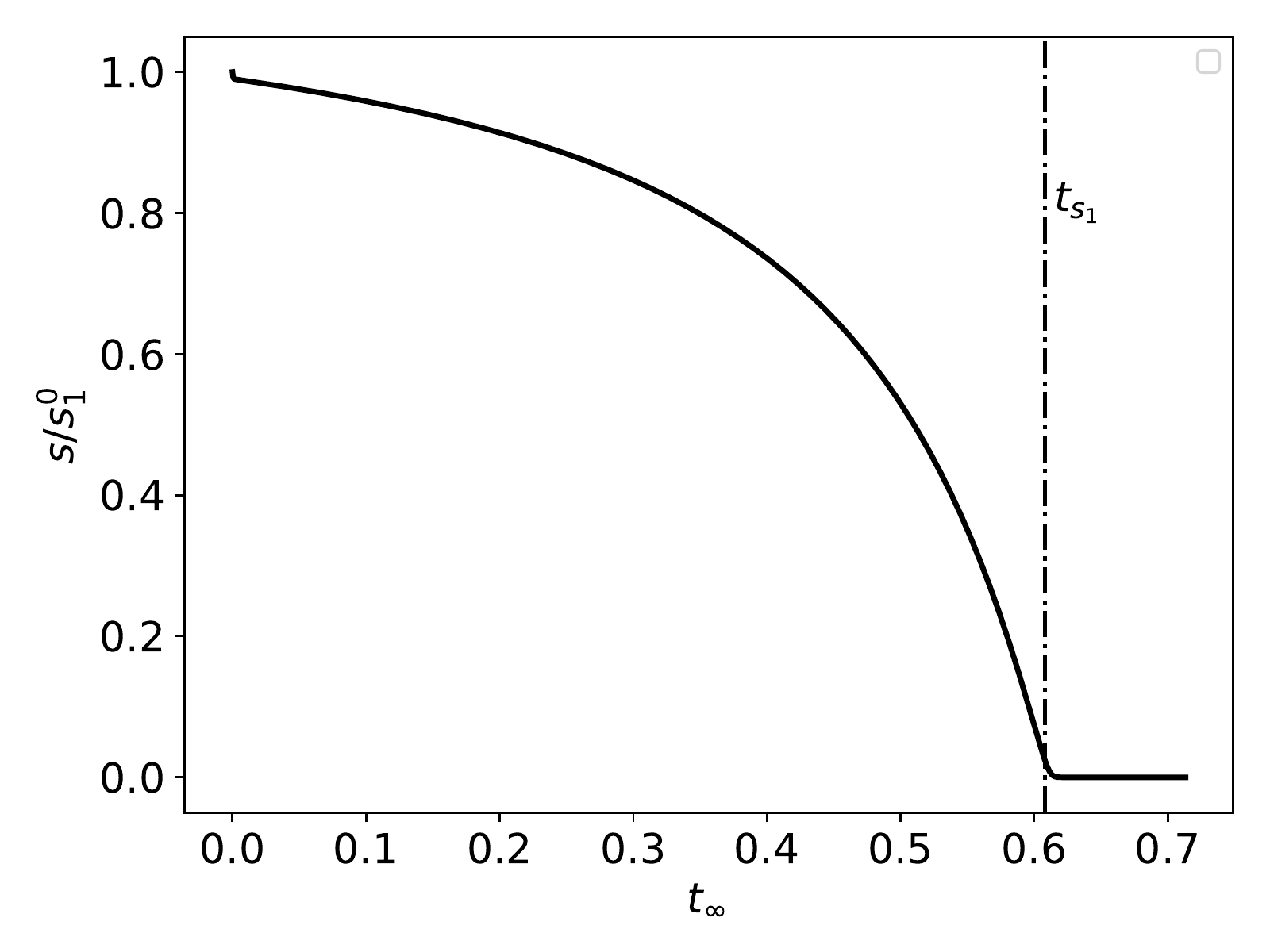}
\caption{\textbf{The graphical illustrations of the completion timescale for 
the Michaelis--Menten reaction mechanism (\ref{eq:react1}).} When $\sigma_1 \gg 1$, 
the reaction is essentially complete when $t=t_{s_1}$. The solid black curve is the
numerical solution to the mass action equations~(\ref{eq:MA01})--(\ref{eq:MA02}) and 
the vertical dashed/dotted line corresponds to $t=t_{s_1}$. The constants (without units) 
used in the numerical simulation are: $e_1^0=1$, $k_1=10$, $k_2=10$, $k_{-1}=1$ and 
$s_1^0=100$. Time has been mapped to the $t_{\infty}$ scale: 
$t_{\infty}(t) = 1-1/\ln[t+\exp(1)]$.}
\label{fig:3}
\end{figure}

The calculus of the Lambert-$W$ function determines the relevant chemical interpretation of 
$t_{s_1}$ as $\sigma_1$ increases. When $t=t_{s_1}$, the substrate concentration is, based 
on the RSA, $K_{M_1}W\left[(1-\ell)\sigma_1\right]$. Furthermore,
\begin{equation}\label{eq:diff}
 W\left[u\right]\ll u, \;\; \text{when} \;\; 1 \ll u,
\end{equation}
and we see from (\ref{eq:diff}) that as the argument ``$u$" gets large, the distance from 
$u$ to $W[u]$ gets greater. Since 
\begin{equation}
s_1(t_{s_1}) = K_{M_1} W[(1-\ell)\sigma_1)],
\end{equation}
it follows from (\ref{eq:diff}) that
\begin{equation}
W[(1-\ell)\sigma_1] \ll (1-\ell)\sigma_1
\end{equation}
as $\sigma_1$ gets large. Thus, for large $\sigma_1$, it holds that
\begin{equation}
s_1(t_{s_1}) \ll (1-\ell)s_1^0,
\end{equation}
in which case we categorize $t_{s_1}$ as a \textit{completion timescale}, since it is 
proportional to the total length of the reaction (see {\sc Figure}~
\ref{fig:3}). 

In the intermediate range, when neither $\sigma_1\ll 1$ or $\sigma_1\gg 1$ holds, $t_{s_1}$ 
is still the appropriate timescale over which a \textit{significant} reduction in
substrate concentration occurs, and in this case we refer to the slow timescale as the 
\textit{depletion timescale}, since it is too long to be a characteristic timescale, 
but too short to be a completion timescale.

\subsection{The QSSA versus the RSA}
How did Segel's work reconciles the work of \citet{HEINEKEN196795} with the observations 
made by \citet{PALSSON1987447}? In a nutshell, \citet{Segel1989} found that over the fast 
timescale the mass action equations scale as
\begin{subequations}
\begin{align}
\dot{\hat{s}}_1 &= \varepsilon_1 f(\hat{s}_1,\hat{c}_1),\\
\dot{\hat{c}}_1 &=  g(\hat{s}_1,\hat{c}_1),
\end{align}
\end{subequations}
and on the slow timescale as
\begin{subequations}
\begin{align}
{\hat{s}_1}' &= \tilde{f}(\hat{s}_1,\hat{c}_1),\\
\varepsilon_2{\hat{c}_1}' &=  \tilde{g}(\hat{s}_1,\hat{c}_1),
\end{align}
\end{subequations}
where $f,g$ denote the right hand sides of (\ref{eq:5}), and $\tilde{f},\tilde{g}$ denotes 
the right hand sides of (\ref{eq:QSSA}). If $\varepsilon_1 \ll 1$, then the depletion of 
substrate over the fast timescale is negligible. However, if $\varepsilon_1 \approx 1$, 
but $\varepsilon_2 \ll 1$, then the QSSA is still valid after a brief transient. The 
distinguishing feature in the case when $\varepsilon_2 \ll \varepsilon_1 \sim 1$ is that 
the depletion of $s_1$ over the initial transient is noticeable \citep{Segel1989}. 

It is straightforward to the show that $\varepsilon_2 < \varepsilon_1$. Consequently, the 
condition for the validity of the RSA, $\varepsilon_1 \ll 1$, ensures the validity of the 
QSSA on the slow timescale. Moreover, since $\varepsilon_1 \ll 1$ guarantees that the 
depletion of $s_1$ minimal over $t_{c_1}$, the qualifier $\varepsilon_1 \ll 1$ ensures
the validity of the QSSA for the entire dynamics of the 
reaction~(\ref{eq:react1}) \citep{Hanson:2008:RSA}. 

\subsection{Matched asymptotics: The composite solution for the time course of the reaction}
\begin{figure}[htb!]
\centering
\includegraphics[width=12cm]{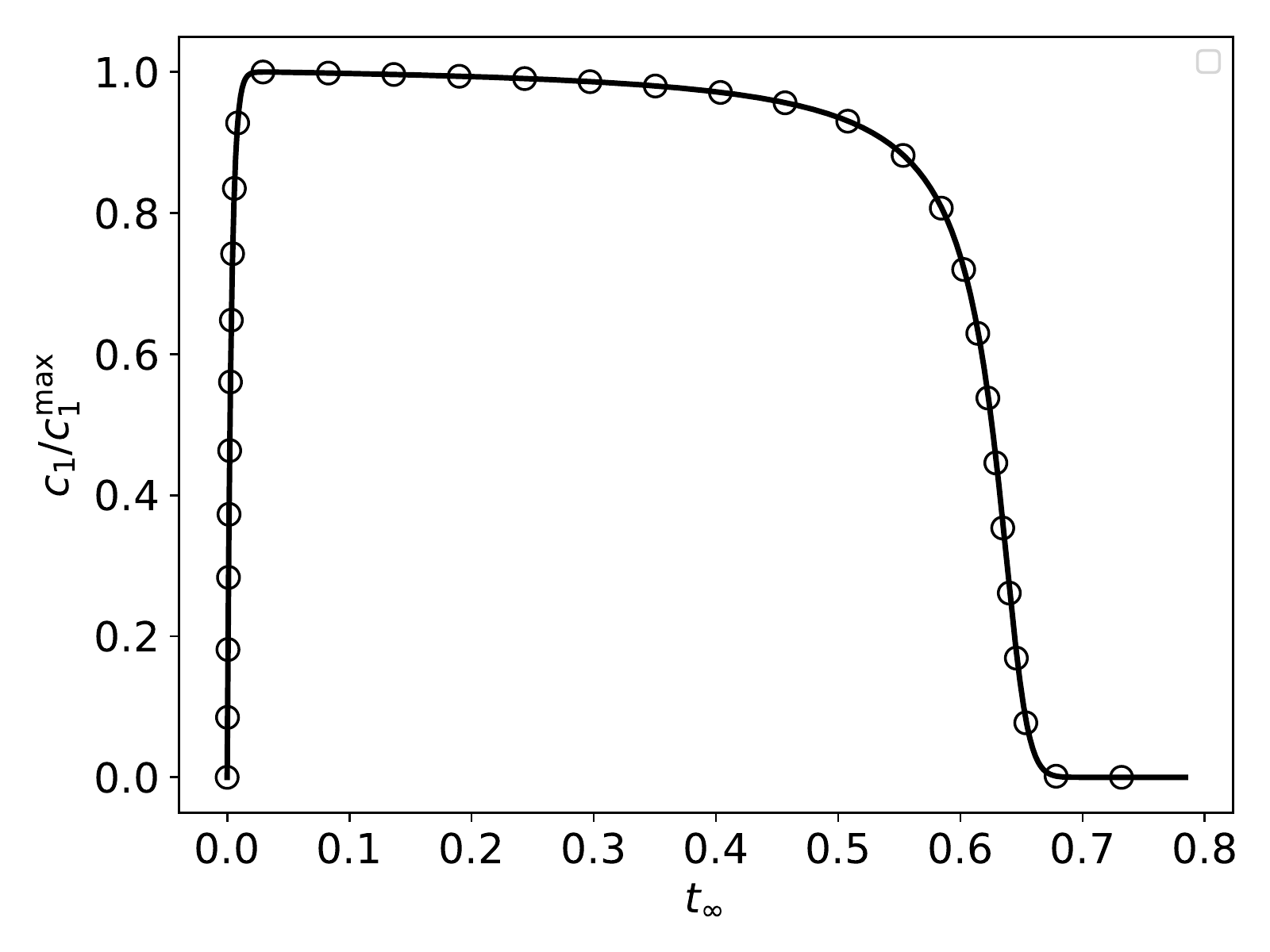}
\caption{\textbf{A graphical comparison of the composite and numerical solutions for the 
time course of the Michaelis--Menten reaction (\ref{eq:react1})}. The solid black curve
is the numerical solution to (\ref{eq:MA01})--(\ref{eq:MA02}). The unfilled circles mark 
the composite solution~(\ref{eq:comp}). The initial concentrations and rate constants used 
in the numerical simulation are: $k_1=1$, $k_2=1$, $k_{-1}=1$, $e_1^0=1$ and $s_1^0=100$ 
(units have been omitted). All approximations have been scaled by their numerically--obtained 
maximum values, and time has been mapped to the $t_{\infty}$ scale: 
$t_{\infty}(t) = 1-1/\ln[t+\exp(1)]$.}
\label{fig:4}
\end{figure}

Expressing the asymptotic solution to (\ref{eq:MA01})--(\ref{eq:MA02}) as,
\begin{subequations}
\begin{align}
&\begin{cases}\label{eq:10}
s_1 \simeq s_1^0,\\
\qquad \qquad \qquad \qquad \quad \qquad \qquad \qquad  t \leq t_{c_1}\\
c_1 \simeq c_1^{\max}\left[1-\exp (-t/t_{c_1})\right],
\end{cases}
\\ \nonumber \\
&\begin{cases}\label{eq:9}
s_1 \simeq K_{M_1} W\left[\sigma_1 \exp(\sigma_1 -\eta_1 t)\right],\\
\qquad \qquad \qquad \qquad \quad \qquad \qquad \qquad t > t_{c_1}\\
c_1 \simeq \cfrac{e_1^0}{K_{M_1}+s_1}s_1,
\end{cases}
\end{align}
\end{subequations}
serves well to convey the fact that the dynamics of the reaction changes depending on 
where a particular time point falls in relation to $t_{c_1}$, and these equations provide 
us with the correct inner and outer solutions that approximate the kinetics under the RSA. 
However, it is well-understood that equations~(\ref{eq:10})--(\ref{eq:9}) are misleading: 
there is a large transition regime surrounding $t_{c_1}$ and, within this transition regime, 
the outer solution~(\ref{eq:9}) does not accurately approximate the solution. Note that the 
presence of a transition regime does not suggest that $t_{c_1}$ is inappropriate timescale.  
In fact, the timescales derived in the previous section are \textit{the} appropriate timescales 
that categorize the fast and slow regimes of the reaction. To see why, and to mitigate the 
effect of the transition region, we construct the \textit{composite solution} for the 
intermediate complex, $c_1^{io}$:
\begin{equation}\label{eq:comp}
c_1^{io} = \cfrac{e_1^0}{K_{M_1}+s_1}s_1 - c_1^{\max}\exp(-t/t_{c_1}).
\end{equation}
The composite solution provides a uniform asymptotic solution that is valid for all time.
Furthermore, the accuracy of (\ref{eq:comp}) indicates that $t_{c_1}$ and $t_{s_1}$ quantify 
the appropriate temporal length scales of the initial fast transient and quasi-steady-state 
regime (see {\sc Figure}~\ref{fig:4}).

\subsection{The characteristic timescale is not a matching timescale}
From a theoretical point of view, the composite solution has little advantage over the 
numerical solution in terms of estimating \textit{when} the transition to the 
quasi-steady-state phase occurs. We will refer to the time at which the transition to 
QSS occurs as a \textit{matching timescale}, and a rough candidate for a matching timescale 
is $t_{c_1}$. The caveat with utilizing $t_{c_1}$ as a matching timescale is that $t_{c_1}$ 
is a characteristic timescale, and hence will only provide characteristic (as opposed to 
limiting) values of the concentrations within a given regime. To clearly illustrate 
why $t_{c_1}$ fails to be an adequate matching timescale requires a phase--plane analysis 
of the mass action equations~(\ref{eq:MA01})--(\ref{eq:MA02}). After the initial buildup 
of the intermediate, $c_1$, the phase--plane trajectory is asymptotic to a slow manifold,
$\mathcal{M}_{\varepsilon}$. The slow manifold is invariant, and is at a
$\mathcal{O}(\varepsilon_2)$-distance from the $c_1$-nullcline, $\mathcal{M}_0$:
\begin{equation}
\mathcal{M}_0 = \bigg\{(s_1,c_1): c_1 - \cfrac{e_1^0}{K_{M_1}+s_1}s_1=0 \bigg\}.
\end{equation}
The outer solution, (\ref{eq:10}), is valid once the trajectory is \textit{extremely close} 
to the slow manifold, which implies $c_1$ should be near its maximum value at the onset of 
the outer solution validity. The complex reaches its maximum value once the trajectory 
reaches $\mathcal{M}_0$. However, when $t=t_{c_1}$, the concentration of the complex is far 
enough away from its maximum value to render the outer solution invalid:
\begin{equation}
c_1(t_{c_1}) \approx \ell c_1^{\max} < c_1^{\max}.
\end{equation}
Thus, $c_1(t_{c_1})\not\in \mathcal{M}_0$, and therefore the trajectory is not quite close 
enough to $\mathcal{M}_{\varepsilon}$ to justify (\ref{eq:10}) as an asymptotic solution 
(see {\sc Figure}~\ref{fig:2}).

\begin{figure}[htb!]
\centering
\includegraphics[width=12cm]{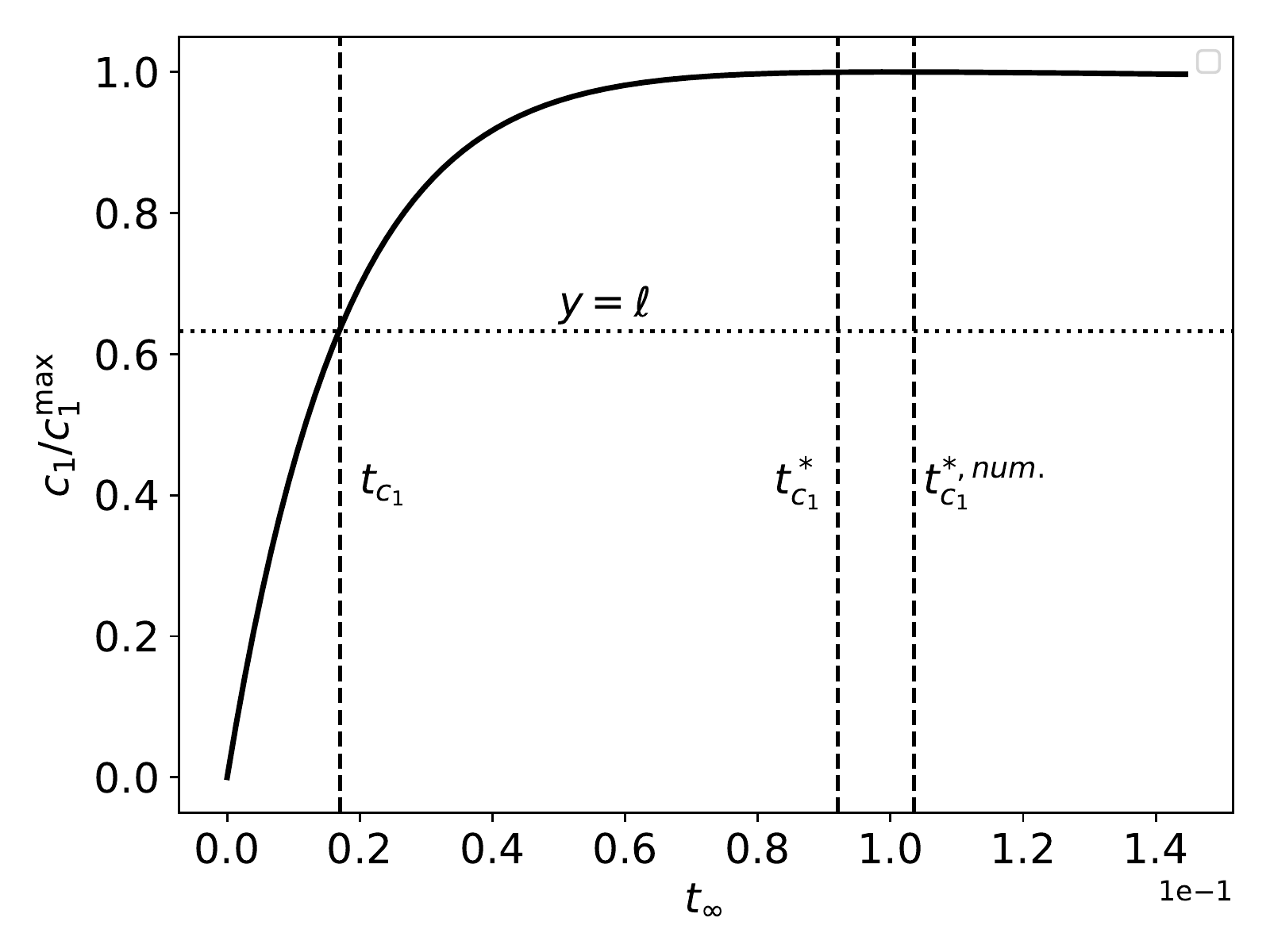}
\caption{\textbf{The validity of $t_{c_1}^{\ast}$ and a graphical representation of its 
comparison with $t_{c_1}$ for the Michaelis--Menten reaction mechanism (\ref{eq:react1})}. 
The solid black curve is the numerically-computed solution to (\ref{eq:MA01})--(\ref{eq:MA02}). 
The left-most dashed vertical curve is corresponds to $t_{c_1}$, and the middle dashed 
vertical curve corresponds to the estimated value $t_{c_1}^{\ast}=-t_{c_1}\ln \varepsilon_2$. 
The dashed vertical line corresponds to the numerically-computed $t_{c_2}^{\ast}$, which is 
labeled as $t_{c_1}^{\ast,num}$ in the figure. Notice that $t_{c_1}^{\ast.}$ provides a much 
better estimate of the time it takes $c_1$ to reach its maximum than $t_{c_1}$. The initial
concentrations and rate constants used in the numerical simulation are: $k_1=0.1$, $k_2=10$,
$k_{-1}=1$, $e_1^0=1$ and $s_1^0=100$ (units have been omitted). Time has been mapped to the
$t_{\infty}$ scale: $t_{\infty}(t) = 1-1/\ln[t+\exp(1)]$. Note that the mass action
equations have only been integrated from $t=0$ to $t\approx t_{c_1}^{\ast}$ for clarity.} 
\label{fig:5}
\end{figure}

A more accurate estimate of the actual time it takes $c_1$ to reach its maximum concentration 
(we will denote this timescale as $t_{c_1}^{\ast}$) can be obtained by either: (i) solving 
the mass action equations exactly or, (ii) by means of an asymptotic approximation. Employing 
strategy (i) is difficult due to the nonlinearity of the equations; strategy (ii) tends to be 
more straightforward to implement. If utilize (ii), we immediately meet with an obvious conundrum 
if we try to estimate $t_{c_1}^{\ast}$ directly from (\ref{eq:10}) or (\ref{eq:9}): (\ref{eq:10}) 
predicts that it will take an infinite amount of time for $c_1$ to reach $c_1^{\max}$, 
while (\ref{eq:9}) predicts $t_{c_1}^{\ast} = 0$. To work around this, we look for an 
asymptotic estimate to $t_{c_1}^{\ast}$. Staring with the inner solution,
\begin{equation}\label{eq:tau}
c_1(\tau)=c_1^{\max}\left[1-\exp{\displaystyle(-\tau)}\right]
\end{equation}
we rewrite (\ref{eq:tau}) in terms of the slow variable, $T=t/t_{s_1}$:
\begin{equation}\label{eq:tauT}
c_1(T)=c_1^{\max}\left[1-\exp{\displaystyle(-T/\varepsilon_2)}\right].
\end{equation}
By inspection of (\ref{eq:tauT}), we see that $c_1$ \textit{should} be in an $\mathcal{O}(\varepsilon_2)$-neighborhood of the slow manifold when
\begin{equation}\label{eq:TT}
T = \varepsilon_2 |\ln \varepsilon_2|.
\end{equation}
Next, since $T=t/t_{s_1}$, we solve for $t$ in (\ref{eq:TT}) to obtain an asymptotic estimate 
on $t_{c_1}^{\ast}$:
\begin{equation}\label{eq:c1ast}
t_{c_1}^{\ast}\simeq-t_{c_1}\ln \varepsilon_2.
\end{equation}
The timescale (\ref{eq:c1ast}) is the matching timescale, although various authors refer 
to any timescale of order $\varepsilon |\ln \varepsilon|$ as simply 
\textit{a slow time} \citep{kuehn2015multiple}. While not exact, the 
approximation~(\ref{eq:c1ast}) provides a useful estimate of the 
time to transition from transient to quasi-steady-state kinetics for the single-enzyme, single-substrate MM reaction mechanism (see {\sc Figure}~\ref{fig:5}).

As a final remark, we note that the asymptotic approximation (\ref{eq:c1ast}) is not without 
a more rigorous justification. So far, we have been able to estimate matching timescales by 
directly calculating them from the ``inner" or transient solution; the direct method is possible 
because we have closed-form solutions comprised of exponential functions. However, for a generic 
fast/slow dynamical system of the form
\begin{subequations}
\begin{align}
\dot{x} &= f(x,y),\label{eq:slow}\\
\varepsilon \dot{y} &= g(x,y)\label{eq:fast},
\end{align}
\end{subequations}
the equation $\dot{y}=g(x_0,y)$ may not be linear, and a closed-form solution may not be possible. 
However, it is a well-known fact, stated in both textbooks \citep{kuehn2015multiple} and 
literature \citep{Klonowski1983}, that the time necessary for the fast-variable to reach QSS is 
generally $\mathcal{O}(\varepsilon |\ln \varepsilon|)$. This result is due to the work 
of \citet{tikhonov1952}, who studied the convergence of the solution to the perturbed 
system (\ref{eq:slow})--(\ref{eq:fast}) to the solution of the degenerate 
system, (\ref{eq:slowD})--(\ref{eq:fastD}):
\begin{subequations}
\begin{align}
\dot{x} &= f(x,y),\label{eq:slowD}\\
0 &= g(x,y)\label{eq:fastD}.
\end{align}
\end{subequations}
The work of Tikhonov is summarized as follows: First, (\ref{eq:fastD}) defines a corresponding 
slow manifold of the form $y=h(x)$, where $g(x,h(x))=0$. Next, let $D$ be the domain over which 
$h:D\to \mathbb{R}^n$ is continuous. If $g$ and $f$ are sufficiently smooth, then the 
following theorem provides a more general technical justification for (\ref{eq:c1ast}):

\begin{theorem}\label{eq:theorem1}{Convergence towards the slow manifold:}
Suppose the system (\ref{eq:slow})--(\ref{eq:fast}) has an associated slow manifold, 
$\mathcal{M}_0=\{(x,y): y=h(x) \;\&\; x\in D \}$, that is uniformly asymptotically stable. If 
$f$, $g$ and their first two derivatives are uniformly bounded in a neighborhood ``$N$" 
of $\mathcal{M}_0$, then there are positive constants $\varepsilon_0$, $b_0$, $b_1$, $\Lambda$, 
and $M$ such that for any initial condition $(x_0,y_0)\in N$ such that 
$||y_0 - h(x_0)|| \leq b_0$, and any $\varepsilon$ such that $0 < \varepsilon < \varepsilon_0$, 
the bound 
\begin{equation}\label{eq:bound}
||y(t) - h(x(t))|| \leq M||y_0 - h(x_0)||\exp\left[\displaystyle -\Lambda t/\varepsilon\right] + b_1\varepsilon,
\end{equation}
holds provided $x(t) \in D$. 
\end{theorem}
Notice the slow manifold utilized in the theorem is not defined to be \textit{invariant}. 
In fact, $\mathcal{M}_0$ is the nullcline associated with the fast variable, $y$, and is 
formally referred to as \textit{the critical manifold}. The non-invariant slow manifold 
employed in Theorem~(\ref{eq:theorem1}) arises from the original form of the theorem 
introduced by \citet{tikhonov1952}. The specific form of Theorem (\ref{eq:theorem1}) is 
taken directly from \citet{Berglund}, but originally introduced by \citet{Gradshtein1953}.
\citet{FENICHEL197953} later extended slow/fast theory by demonstrating that there exists 
an \textit{invariant} slow manifold that is present in the phase-space of the system when 
$\varepsilon$ is sufficiently small but non-zero.

What the bound specifically tells us is that if $t = \varepsilon |\ln \varepsilon|$, then
\begin{equation}\label{eq:boundless}
||y(t) - h(x(t))|| \leq M||y_0 - h(x_0)||\varepsilon^{\Lambda} + b_1\varepsilon,
\end{equation}
and thus the phase-plane trajectory should be at a distance that is $\mathcal{O}(\varepsilon)$ 
from $\mathcal{M}_0$ once $t = \varepsilon |\ln \varepsilon|$ \citep[see][for details]{Berglund}. 
In a fast/slow system of the form (\ref{eq:slow})--(\ref{eq:fast}), the small parameter 
$\varepsilon$ is proportional to the ratio of the fast and slow timescales. Moreover, the
system~(\ref{eq:slow})--(\ref{eq:fast}) is assumed to be dimensionless. Thus, if we apply
Theorem~(\ref{eq:theorem1}) to
\begin{equation}\label{eq:QSSA1}
\begin{aligned}
\cfrac{d\hat{s}_1}{dT}&=(1+\kappa_1)(1+\sigma_1)\left [-\hat{s}_1 + (1-\beta_1)\hat{c}_1\hat{s}_1 + \beta_1\alpha_1\hat{c}_1\right],\\
\varepsilon_2 \cfrac{d\hat{c}_1}{dT}&= \hat{s}_1 - (1-\beta_1)\hat{c}_1\hat{s}_1 - \beta_1\hat{c}_1, 
\end{aligned}
\end{equation}
then the phase--plane trajectory should be $\mathcal{O}(\varepsilon_2)$ from the $c_1$-nullcline 
when $T = \varepsilon_2 |\ln \varepsilon_2|$. Consequently, since $T = t/t_{s_1}$, we obtain
\begin{equation}
t = t_{s_1}\cdot\varepsilon_2 |\ln \varepsilon_2| = -t_{c_1}\ln \varepsilon_2 \approx t_{c_1}^{\ast}
\end{equation}
as the asymptotic time required for $c_1$ to reach its maximum value.

The calculation of the matching timescale is more than just an exercise: there is chemical 
utility in computing $t_{c_1}^{\ast}$. Specifically, it indicates approximately when the 
rate of product formation reaches its maximum quasi-steady-state production:
\begin{equation}
\max \dot{p} \simeq \dot{p}(t_{c_1}^{\ast}).
\end{equation}
Thus, the matching timescale is a very good indication of how long it takes before the 
product formation rate reaches its maximum value, and when the reaction can be assumed to 
be in a quasi-steady-state phase.

\section{The auxiliary enzyme reaction mechanism}
We now consider the more complicated case of the auxiliary enzyme reaction 
mechanism \citep{Eilertsen:2018:KAS}. The mechanism is composed of two reactions: 
a primary reaction (\ref{eq:react2}) that produces a substrate, $S_2$, that is synthesized 
in a catalytic step:
\begin{align} \label{eq:react2}
E_1 + S_1&   \begin{array}{c} {\scriptstyle k_1}\\
                       \rightleftharpoons\\
                       {\scriptstyle k_{-1}}
        \end{array} C_1 
        \begin{array}{c}
        {\scriptstyle k_2}\\
        \rightarrow \\
        {}
        \end{array} E_1 + S_2,
\end{align}
and a secondary reaction, (\ref{eq:react4}), where $S_2$ binds with the auxiliary enzyme 
``$E_2$" and releases the final product, $P$: 
\begin{align} \label{eq:react4}
E_2 + S_2&   \begin{array}{c} {\scriptstyle k_3}\\
                       \rightleftharpoons\\
                       {\scriptstyle k_{-3}}
        \end{array} C_2 
        \begin{array}{c}
        {\scriptstyle k_4}\\
        \rightarrow \\
        {}
        \end{array} E_2 + P .
\end{align}
The complete set of mass action equations that model the kinetics of the complete reaction 
mechanism (\ref{eq:react2})--(\ref{eq:react4}) are
\begin{subequations}\label{eq:MA}
\begin{align}
\dot{s}_1 &= -k_1(e_1^0-c_1)s_1 + k_{-1}c_1\label{eq:MA1},\\
\dot{c}_1 &= k_1(e_1^0-c_1)s_1 -(k_{-1}+k_2)c_1\label{eq:MA2},\\
\dot{s}_2 &= -k_3(e_2^0-c_2)s_2 + k_{-3}c_2 + k_2c_1\label{eq:MA3},\\
\dot{c}_2 &= k_3(e_2^0-c_2)s_2 -(k_{-3}+k_4)c_2\label{eq:MA4},
\end{align}
\end{subequations}
where $s_1$ and $s_2$ denote the respective concentrations of the substrates $S_1$ and $S_2$, 
$c_1$ and $c_2$ denote the concentrations of the complexes $C_1$ and $C_2$, and $e_1^0$ and 
$e_2^0$ denote the initial concentrations of the primary and auxiliary enzymes, $E_1$ and 
$E_2$. $k_3$ and $k_{-3}$ are rate constants, and $k_4$ is the catalytic constant of 
the secondary reaction. We define the initial conditions for the secondary reaction 
as $(s_2,c_2)(t=0)=(0,0)$.

In forthcoming analysis, we will assume that the primary reaction obeys the RSA (i.e., 
$\varepsilon_1 \ll 1$). Additionally, we will make the assumption that $k_2 \lesssim k_4$, 
and that the initial auxiliary enzyme concentration is larger than $e_1^0$ (i.e., 
$e_1^0=1$, $e_2^0 \gg 1$). We also compute matching timescales that yield a reliable estimate of the time it takes $s_2$ and $c_2$ to reach QSS. Moreover, a new timescale called the \textit{lag time} will be introduced. The lag time corresponds to the time it takes $\dot{p}$ to reach its maximum value, and we will show that it corresponds to a specific matching timescale. Thus, not only do matching timescales provide estimates for the time it takes a specific species to reach QSS, they also, in the context of auxiliary reactions, provide an approximation of the time it takes before the complete reaction begins generating product at a maximal rate. 

\subsection{The study of phase--plane geometry of the auxiliary enzyme reaction mechanisms 
permit a heuristic estimation of characteristic timescales}
\begin{figure}[htb!]
\centering
\includegraphics[width=12cm]{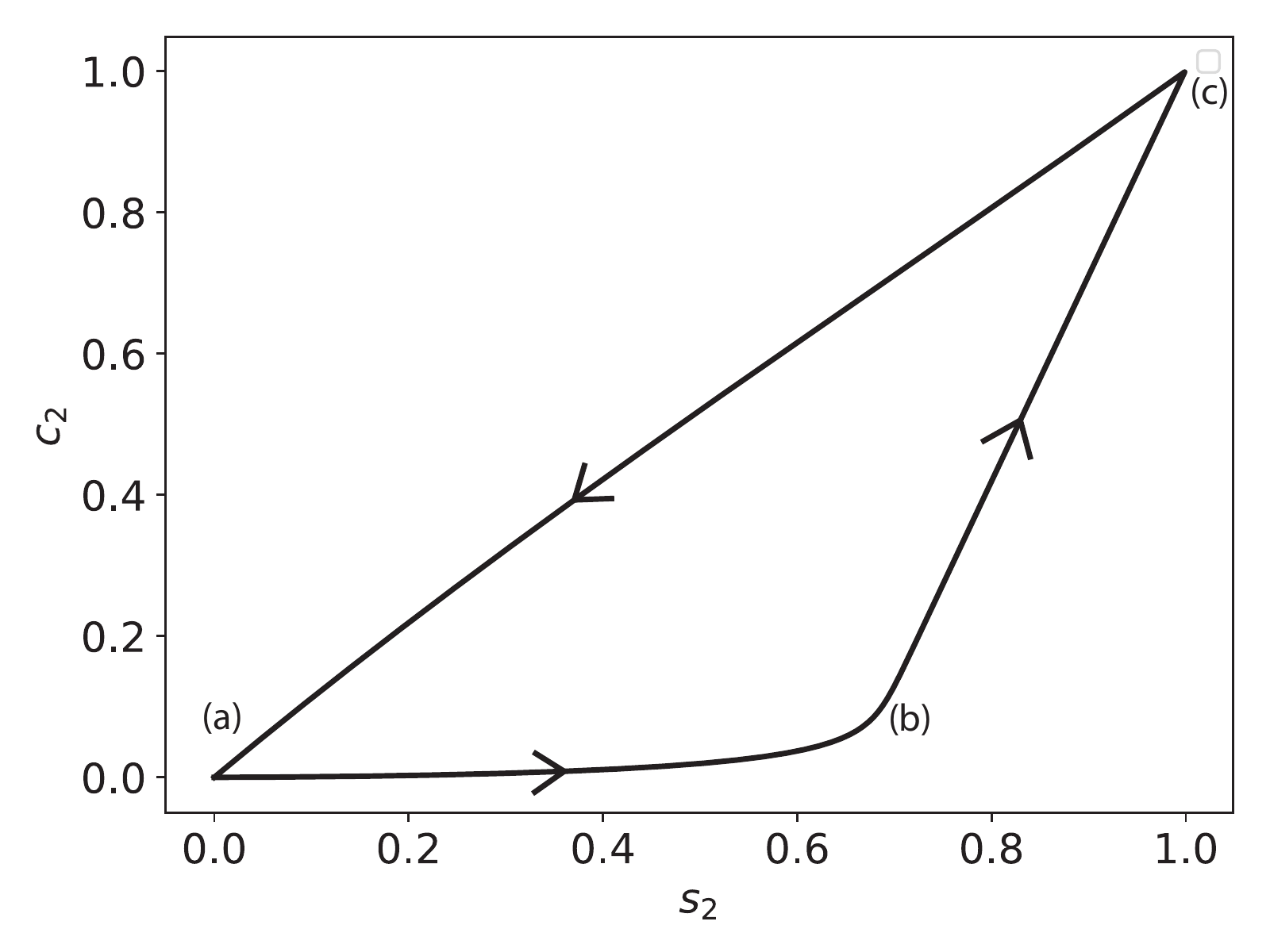}
\caption{\textbf{The phase--plane portrait of the mass action trajectory for the auxiliary 
reaction mechanism (\ref{eq:react2})--(\ref{eq:react4})}. The solid black curve is the 
numerically-computed solution to (\ref{eq:MA1})--(\ref{eq:MA4}). The initial concentrations 
and rate constants used in the numerical simulation are: $k_1=1$, $k_2=1$, $k_{-1}=1$, $e_1^0=1$, 
$e_2^0=100$, $k_{-3}=1$, $k_{3}=1$, $k_4=2$ and $s_1^0=100$ (units have been omitted). $s_2$ and 
$c_2$ have been scaled by their numerically--obtained maximum values.}
\label{fig:6}
\end{figure}

Perhaps the most intuitive way to derive the relevant characteristic timescales of
(\ref{eq:react2})--(\ref{eq:react4}) is to get a qualitative understanding of what 
a typical phase-plane trajectory looks like in the $c_2$--$s_2$ plane. Numerical 
simulations suggest that the phase--plane trajectory is almost ``triangular" in certain 
parameter ranges (see {\sc Figure}~\ref{fig:6}) and,
based on the appearance of the phase--plane trajectory, there seems to be at least three 
distinct timescales:
\begin{description}
\item[$\bullet$] The scale on which the trajectory travels from (a) to (b). We will 
denote this timescale as $t_{s_2}$.
\item[$\bullet$] The scale on which the trajectory travels from (b) to (c). We will 
denote this timescale as $t_{c_2}$.
\item[$\bullet$] The scale on which the trajectory travels from (c) back to (a). We will 
denote this timescale as $t_p$.
\end{description}

The logical step that follows will be to make some initial \textit{a priori} assumptions 
about the ordering of all the timescales involved in the reaction. For the sake of simplicity, 
let us initially assume that $t_{c_1}\ll t_{s_2}, t_{c_2} \ll t_{s_1}$, and that the
\textit{completion} timescale for the secondary reaction is identically $t_{s_1}$. This 
implies that the secondary reaction completes at roughly the same time as the primary 
reaction, and that $t_p \approx t_{s_1}$. Thus, we have eliminated one timescale ($t_p$) 
by imposing the assumption that the secondary reaction is as fast as the primary reaction. 

The next step will be to exploit the presence and geometry of any manifolds (not necessarily 
invariant) that exist within the phase-plane of the secondary reaction. Notice that the 
intersection of the $s_2$-nullcline and $c_2$-nullcline is \textit{time-dependent} since 
the $s_2$-nullcline moves as $c_1$ varies in time. Geometrically, the intersection of the 
nullclines is described by a moving fixed point, $\boldsymbol{x}^{\ast}$,
\begin{equation}
N_{s_2} \cap N_{c_2} \equiv \boldsymbol{x}^{\ast},
\end{equation}
where $N_{s_2}$ denotes the $s_2$-nullcline and $N_{c_2}$ denotes the $c_2$-nullcline. 
Algebraically, the coordinates of $\boldsymbol{x}^{\ast}$, $(s_2^{\ast},c_2^{\ast})$, are
\begin{equation}\label{eq:fixedPT}
s_2^{\ast}= \cfrac{K_{M_2}k_2c_1(t)}{V_2-k_2c_1(t)},\quad c_2^{\ast} = \cfrac{k_2c_1(t)}{k_4},
\end{equation}
where $K_{M_2}$ denotes the Michaelis constant of the secondary reaction
\begin{equation}
K_{M_2} \equiv \cfrac{k_{-3}+k_4}{k_3},
\end{equation}
and $V_2$ denotes the limiting rate of the secondary reaction: $V_2 \equiv k_4e_2^0$. Moreover, 
if the second reaction is as fast as the primary reaction, then the phase-plane geometry 
suggests that the trajectory should not only catch the fixed point $\boldsymbol{x}^{\ast}$, 
but will also approximately \textit{adhere} to $\boldsymbol{x}^{\ast}$ as it descends to the 
origin. If the trajectory adheres to $\boldsymbol{x}^{\ast}$, then 
\begin{equation}\label{eq:Dotp}
\dot{p} =\cfrac{V_2 \cfrac{K_{M_2}k_2c_1}{V_2-k_2c_1}}{K_{M_2}+\cfrac{K_{M_2}k_2c_1}{V_2-k_2c_1}}= k_2c_1,
\end{equation}
and the product formation rate of the secondary reaction has reached its limiting value. 
Notice that by assuming that the secondary reaction is fast enough to virtually adhere to
$\boldsymbol{x}^{\ast}$ implies $V_2 > k_2c_1^{\max}$. Thus, this assumption admits an 
automatic partition of parameter space, and we will only consider regions of parameter 
space within which $V_2 > k_2c_1^{\max}=k_2\varepsilon_1 s_1^0$ holds.

\begin{figure}[hbt!]
\centering
\includegraphics[width=12cm]{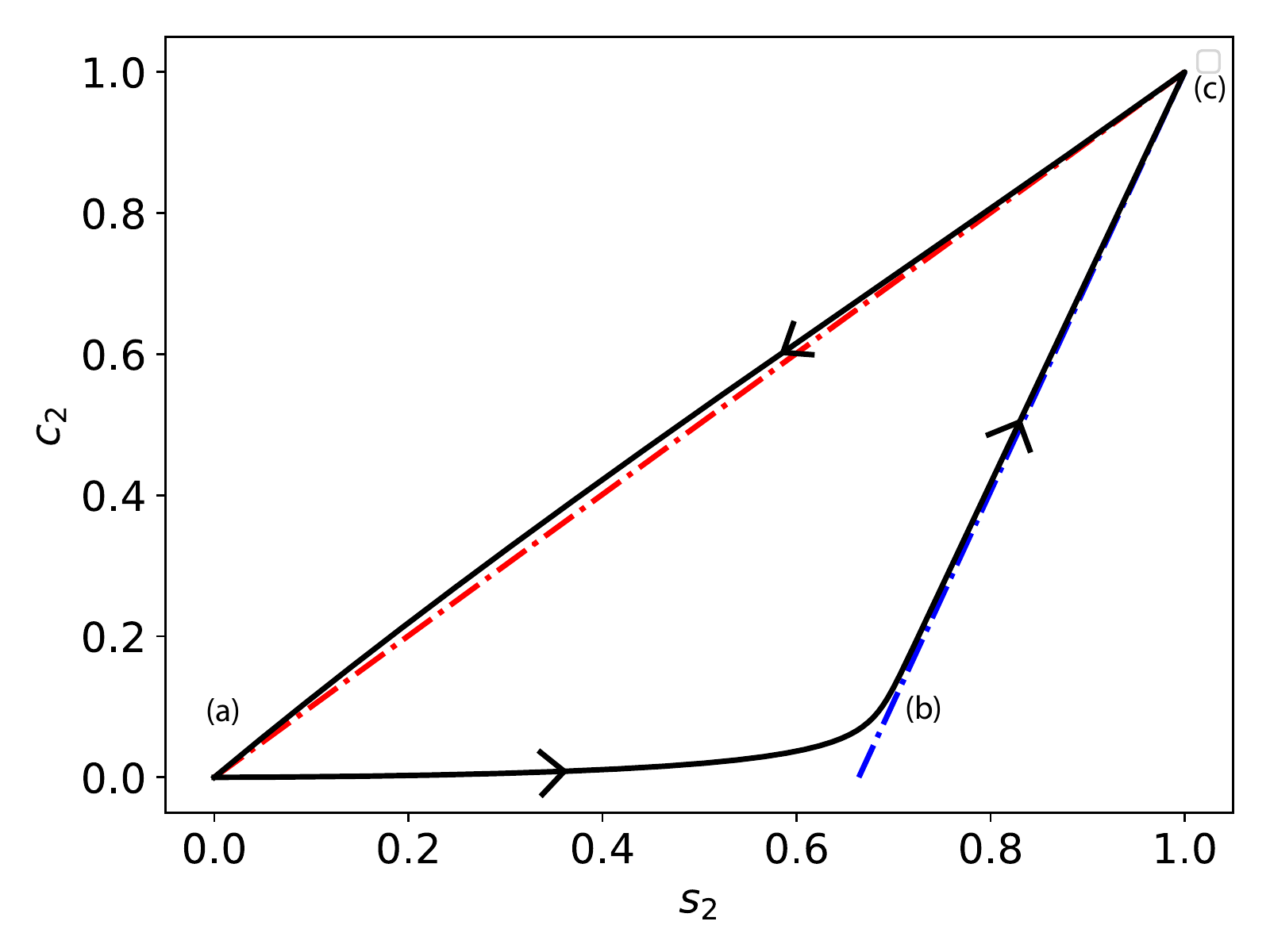}
\caption{\textbf{The $s_2$--$c_2$ phase-plane trajectory (with nullclines) for the auxiliary 
reaction mechanism (\ref{eq:react2})--(\ref{eq:react4})}. The thick black curve is the 
numerically-integrated solutions to the mass action equations (\ref{eq:MA1})--(\ref{eq:MA4}). 
The broken red curve is the $c_2$-nullcline, and the broken blue curve is the fixed 
$s_2$-nullcline ($N_{s_2}^{\max}$, given by (\ref{eq:STATIONARY})). The phase--plane 
trajectory initially moves towards $N_{s_2}^{\max}$, then moves up $N_{s_2}^{\max}$ before 
moving back down the $c_2$-nullcline.  The constants (without units) used in the numerical 
simulation are: $e_1^0=1$, $s_1^0=1000$, $k_1=1$, $k_2=1$, $k_3=1$, $k_{-3}=1$, $k_4=2$, 
$e_2^0=100$ and $k_{-1}=1$. Curves were scaled by their numerically obtained maximum values.}
\label{fig:7}
\end{figure}

Since the position of the $s_2$-nullcline depends on the concentration $c_1$, we want to 
estimate how $c_1$ varies over the course of the reaction. As we are assuming that the primary 
reaction follows the RSA, the phase plane trajectory will follow a slow manifold when 
$t\geq t_{c_1}^{\ast}$. If we know the shape of the slow manifold, then we can get a rough 
idea of how $c_1$ varies throughout the reaction. To do this, we will look at the 
dimensionless equations
\begin{equation}\label{eq:QSSA2}
\begin{aligned}
\cfrac{d\hat{s}_1}{dT}&=(1+\kappa_1)(1+\sigma_1)\left [-\hat{s}_1 + (1-\beta_1)\hat{c}_1\hat{s}_1 + \beta_1\alpha_1\hat{c}_1\right],\\
\varepsilon_2 \cfrac{d\hat{c}_1}{dT}&= \hat{s}_1 - (1-\beta_1)\hat{c}_1\hat{s}_1 - \beta_1\hat{c}_1. 
\end{aligned}
\end{equation}
The zeroth order asymptotic approximation to the slow manifold is  the $\hat{c}_1$-nullcline:
\begin{equation}
\hat{s}_1 -(1-\beta_1)\hat{c}_1\hat{s}_1 - \beta_1\hat{c}_1=0.
\end{equation}
Notice that $\beta_1 \to 0$ as $\sigma_1 \to \infty$; thus, as $\sigma_1 \to \infty$, the 
trajectory that follows the slow manifold will be asymptotic to the curve $\hat{c}_1 = 1$ 
for most of the reaction. Hence, when $\sigma_1 \gg 1$, the concentration of the intermediate 
complex remains near its maximum value, $c_1^{\max}$, for the majority of the reaction, and 
the $s_2$-nullcline will be effectively stationary after the initial buildup of $c_1$. Under 
the assumption that $t_{c_1}$ is the shortest timescale, the initial transient behavior of 
$c_2$ will occur while the $s_2$-nullcline remains fixed. Thus, we look at the phase--plane 
trajectory with the $s_2$-nullcline (with fixed $c_1$) at its stationary value 
(see {\sc Figure}~\ref{fig:7}). Let us denote this manifold as $N_{s_2}^{\max}$:
\begin{equation}\label{eq:STATIONARY}
N_{s_2}^{\max} \equiv \bigg\{(s_2,c_2)\in \mathbb{R}^2:c_2 - \cfrac{k_3e_2^0s_2-k_2c_1^{\max}}{k_3s_2 + k_{-3}}=0\bigg\}.
\end{equation}

Next, we want to exploit the phase-plane geometry in order to estimate critical timescales. 
We will first estimate $t_{s_2}$ by noting that the phase--plane trajectory essentially lies 
along the $s_2$--axis for $t \leq t_{s_2}$. This suggests that
\begin{equation}
\dot{s}_2 \approx -k_3s_2 +k_2c_1, \quad t\leq t_{s_2}
\end{equation}
is a reasonable approximation to (\ref{eq:MA3}). If the initial fast transient of the primary 
reaction is negligibly short, i.e., $t_{c_1}\ll t_{s_2}$, then it is reasonable to assume
\begin{equation}\label{eq:linear1}
\dot{s}_2 \approx -k_3s_2 +k_2c_1^{\max}, \quad t\leq t_{s_2}.
\end{equation}
Since (\ref{eq:linear1}) is linear, its exact solution
\begin{equation}
s_2 \approx s_2^{\lambda}\left[1-\exp(-t/t_{s_2})\right]
\end{equation}
provides two critical estimates: the characteristic timescale, $t_{s_2}$, and an approximate 
maximum value of $s_2$ on the $t_{s_2}$ timescale:
\begin{equation}
t_{s_2} \equiv \cfrac{1}{k_3e_2^0}, \quad s_2 \leq s_2^{\lambda} \equiv \cfrac{k_2c_1^{\max}}{k_3e_2^0}.
\end{equation}
The prediction that $s_2 < s_2^{\max}$ for $t \leq t_{s_2}$ (obtained from the linear equation) 
is in qualitative agreement with the phase-plane trajectory of the numerically-integrated 
equations ({\sc Figure}~\ref{fig:7}). 

Next, to estimate $t_{c_2}$, we note that since the phase--plane trajectory lies close to $N_{s_2}^{\max}$ along its ascension to $c_2^{\max}$, the growth of the intermediate complex 
is approximately
\begin{equation}
\dot{c}_2 \approx -k_4c_2 +k_2c_1^{\max}, \quad t_{s_2} \leq t\leq t_{c_2},
\end{equation}
which admits an analytical solution:
\begin{equation}\label{eq:c2a}
c_2 \approx c_2^{\max}\left[1-\exp(\displaystyle-k_4t)\right].
\end{equation}
Trajectories that follow the $s_2$-nullcline closely are said to be in rapid 
equilibrium \citep{HNsF,Roussel:1991:ASS} or a reverse quasi-steady-state \citep{SCHNELL2000483}.
This is in contrast to trajectories that follow the $c_2$-nullcline, which are said to 
be in a quasi-steady-state phase \citep{Eilertsen:2018:KAS}. From (\ref{eq:c2a}), we have 
two observations: (i) $k_4^{-1}$ is a reasonable estimate of $t_{c_2}$, and (ii) this 
linearized solution predicts $c_2$ will approach $c_2^{\max}$, which is in qualitative 
agreement with the phase--plane trajectory. 

As a concluding remark of this subsection, we note that there are four timescales, 
$t_{c_1},t_{s_2},t_{c_2}$ and $t_{s_1}$, that influence the overall dynamics of the 
coupled reaction. Only two timescales are needed to characterize the dynamics of the 
single-enzyme, single-substrate MM reaction mechanism. Thus, only the ordering of 
two timescales, $t_{c_1}$ and $t_{s_1}$, needs to be considered. In the case of the 
coupled reaction, there are multiple orderings that need to be considered in order 
to fully comprehend the dynamics. In the immediate subsections that follow, we will 
analyze the dynamics with respect to the orderings: 
$t_{c_1}\ll t_{s_2}\ll t_{c_2} \ll t_{s_1}$ and $t_{s_2}, t_{c_2} \ll t_{c_1} \ll t_{s_1}$. 
Both analyses of these orderings will be made under the assumption that $e_2^0$ is 
large with respect to $e_1^0$.

\subsection{Scaling analysis: $t_{c_1} \ll t_{s_2} \ll t_{c_2}\ll t_{s_2}$}
Although we now have estimates for the timescales $t_{s_2}$ and $t_{c_2}$, it is important 
to remember that these timescales were obtained under the assumption that $c_1$ is the 
fastest variable (i.e., $c_1$ reaches its maximum before any other variable). We must 
now: (i) determine the appropriate conditions under which approximate adhesion to 
$\boldsymbol{x}^*$ is possible, and (ii) determine the onset of validity 
for (\ref{eq:Dotp}). We begin by scaling the mass action equations. Introducing 
the additional scaled concentrations
\begin{equation}
\hat{s}_2 = s_2/s_2^{\max}, \quad \hat{c}_2 = c_2/c_2^{\max}
\end{equation}
into equations (\ref{eq:MA3})--(\ref{eq:MA4}) admits the dimensionless form:
\begin{subequations}
\begin{align}
\cfrac{d\hat{s}_1}{dT}&=(1+\kappa_1)(1+\sigma_1)\left [-\hat{s}_1 + (1-\beta_1)\hat{c}_1\hat{s}_1 + \beta_1\alpha_1\hat{c}_1\right],\label{eq:scaled0}\\
\varepsilon_2 \cfrac{d\hat{c}_1}{dT}&= \hat{s}_1 - (1-\beta_1)\hat{c}_1\hat{s}_1 - \beta_1\hat{c}_1\label{eq:scaled1}\\
\mu_1\cfrac{d \hat{s}_2}{dT} &=-\hat{s}_2 + (1-\beta_2)\hat{s}_2\hat{s}_2 + \beta_2\alpha_2\hat{c}_2 + r_S \mu_1 \hat{c}_1,\label{eq:scaledA}\\
\mu_2\cfrac{d \hat{c}_2}{dT} &=(1+\kappa_2)(1+\sigma_2)\left[(\hat{s}_2 - (1-\beta_2)\hat{c}_2\hat{s}_2-\beta_2 \hat{c}_2\right]\label{eq:scaledB}.
\end{align}
\end{subequations}
The dimensionless parameters $\kappa_2$, $\sigma_2$, and $r_S$, introduced in 
(\ref{eq:scaled0})--(\ref{eq:scaledB}), are
\begin{equation}
\alpha_2 \equiv\cfrac{\kappa_2}{1+\kappa_2}, \quad \beta_2 \equiv \cfrac{1}{1+\sigma_2}, \quad \kappa_2 \equiv \cfrac{k_{-3}}{k_4}, \quad \sigma_2 \equiv \cfrac{s_2^{\max}}{K_{M_2}}, \quad r_S \equiv \cfrac{s_1^0}{s_2^{\max}}.
\end{equation}
The remaining parameters, $\mu_1$ and $\mu_2$, are the ratios of the secondary reaction 
timescales to the primary reaction substrate timescale:
\begin{equation}\label{eq:Mus}
\mu_1 \equiv \cfrac{t_{s_2}}{t_{s_1}}, \quad \mu_2 \equiv \cfrac{t_{c_2}}{t_{s_1}}.
\end{equation}
It follows from (\ref{eq:Mus}) that if $\{\varepsilon_2, \mu_1, \mu_2\} \ll 1$, then 
the dynamics of (\ref{eq:MA1})--(\ref{eq:MA4}) consist of one slow variable, $s_1$, and 
three fast variables: $c_1$, $s_2$ and $c_2$. The designation of $s_1$ as a slow variable 
and $c_1$, $s_2$ and $c_2$ as fast variables implies that after an initial fast transient, 
the phase--plane trajectory is asymptotic to the intersecting nullclines:
\begin{subequations}
\begin{align}
s_2 &\simeq \cfrac{K_{M_1}}{V_2-k_2c_1}k_2c_1,\label{eq:scaledC}\\
c_2 &\simeq \cfrac{k_2c_1}{k_4}\label{eq:scaledD}.
\end{align}
\end{subequations}
After the initial fast transient of the primary reaction, $k_2c_1$ is asymptotic to
\begin{equation}
k_2c_1 \simeq \cfrac{V_1}{K_{M_1}+s_1}s_1 \equiv -\dot{s}_1^{\varepsilon},
\end{equation}
and thus $c_1$, $s_2$ and $c_2$ are, in the asymptotic limit, explicitly dependent on 
$s_1$ only.

The above approximations, (\ref{eq:scaledC})--(\ref{eq:scaledD}), confirm the hypothesis 
that the phase-plane trajectory follows the intersection of the $s_2$- and $c_2$-nullclines 
as long as the secondary reaction is fast (i.e., $\mu_1,\mu_2 \ll 1$). The additional 
assumption made in the derivation of $t_{s_2}$ and $t_{c_2}$ was that $t_{c_1}$ is the
\textit{shortest} timescale, and that there is no significant formation of $s_2$ or $c_2$ 
for $0 \leq t \leq t_{c_1}$.  To assess the validity of this assumption, we 
rescale (\ref{eq:MA3})--(\ref{eq:MA4}) with respect to $\tau = t/t_{c_1}$:
\begin{subequations}
\begin{align}
\cfrac{d\hat{s}_1}{d\tau}&=\varepsilon_1\left [-\hat{s}_1 + (1-\beta_1)\hat{c}_1\hat{s}_1 + \beta_1\alpha_1\hat{c}_1\right],\label{eq:scaledE0}\\
\cfrac{d\hat{c}_1}{d\tau}&= \left[\hat{s}_1 - (1-\beta_1)\hat{c}_1\hat{s}_1 - \beta_1\hat{c}_1\right]\label{eq:scaledE1}\\
\cfrac{d \hat{s}_2}{d\tau} &=\lambda_1\left[-\hat{s}_2 + (1-\beta_2)\hat{c}_2\hat{s}_2 + \beta_2\alpha_2\hat{c}_2\right] + r_S \varepsilon_2 \hat{c}_1,\label{eq:scaledE}\\
\cfrac{d \hat{c}_2}{d\tau} &=\lambda_2(1+\kappa_2)(1+\sigma_2)\left[\hat{s}_2 - (1-\beta_2)\hat{c}_2\hat{s}_2-\beta_2 \hat{c}_2\right]\label{eq:scaledF}.
\end{align}
\end{subequations}
The parameters that emerge from scaling, $\lambda_1$ and $\lambda_2$, are the ratios we 
need in order to calculate the time that transpires before (\ref{eq:scaledC})--(\ref{eq:scaledD})
become valid approximations:
\begin{equation}
\lambda_1 = \cfrac{t_{c_1}}{t_{s_2}}, \quad \lambda_2 = \cfrac{t_{c_1}}{t_{c_2}}.
\end{equation}
It is straightforward to show that the term ``$r_S \varepsilon_2$" in (\ref{eq:scaledE}) 
is bounded above,
\begin{equation}
r_S \varepsilon_2 < \lambda_1,
\end{equation}
and therefore $s_2$ is slow on $t_{c_1}$ when $\lambda_1 \ll 1$. In addition, 
(\ref{eq:scaledF}) implies that $\lambda_2(1+\kappa_2)(1+\sigma_2)\ll 1$ if $c_2$ is 
to be \textit{slow} over $t_{c_1}$. While it is certainly true that
$\lambda_2(1+\kappa_2)(1+\sigma_2)\ll 1$ is sufficient for $c_2$ to be slow, it is 
not necessary, given that $s_2 \simeq 0$ for $t \leq t_{c_1}$. 

Piecing together the results obtained from the scaling analysis, we obtain
\begin{subequations}
\begin{align}
s_1 &\simeq s_1^0,\\
c_1 &\simeq c_1^{\max}\left[1-\exp(-t/t_{c_1})\right],\\
s_2 &\simeq 0,\\
c_2 &\simeq 0,\\
\end{align}
\end{subequations}
for $t \lesssim t_{c_1}$.

Moving forward, the next ``fastest" timescale in our imposed ordering is $t_{s_2}$. We 
note that in addition to $c_2$ scaling as a slow variable over $t_{c_1}$, the phase-plane 
trajectory indicates that $c_2$ will also be slow over $t_{s_2}$. Thus, we rescale the 
complete set of mass action equations with respect to 
$\bar{T}=t/t_{s_2}, \tilde{s}_2 = s_2/s_2^{\lambda}$ and
$\tilde{c}_2=c_2/e_2^0s_2^{\lambda}/(K_{M_2}+s_2^{\lambda}$):
\begin{subequations}
\begin{align}
\cfrac{d\hat{s}_1}{d\bar{T}}&=\mu_1(1+\kappa_1)(1+\sigma_1)\left [-\hat{s}_1 + (1-\beta_1)\hat{c}_1\hat{s}_1 + \beta_1\alpha_1\hat{c}_1\right],\label{eq:scaledE00}\\
\varepsilon_1\cfrac{d\hat{c}_1}{d\bar{T}}&= \mu_1(1+\kappa_1)(1+\sigma_1)\left[\hat{s}_1 - (1-\beta_1)\hat{c}_1\hat{s}_1 - \beta_1\hat{c}_1\right]\label{eq:scaledE10}\\
\cfrac{d \tilde{s}_2}{d\bar{T}} &=-\tilde{s}_2 + (1-\tilde{\beta}_2)\tilde{c}_2\tilde{s}_2 + \tilde{\beta}_2\alpha_2\tilde{c}_2 + \hat{c}_1,\label{eq:scaledE2}\\
\cfrac{d \tilde{c}_2}{d\bar{T}} &=\nu(1+\kappa_2)(1+\tilde{\sigma}_2)\left[\tilde{s}_2 - (1-\tilde{\beta}_2)\tilde{c}_2\tilde{s}_2-\tilde{\beta}_2 \tilde{c}_2\right]\label{eq:scaledF2}.
\end{align}
\end{subequations}
In (\ref{eq:scaledE2}), the dimensionless parameters $\tilde{\sigma}_2$ and 
$\tilde{\beta}_2$ are given by:
\begin{equation}
\tilde{\sigma}_2\equiv \cfrac{s_2^{\lambda}}{K_{M_2}},\quad \tilde{\beta}_2 = \cfrac{1}{1+\tilde{\sigma}_2}
\end{equation}
Consequently, the production of $s_2$ will be significant on $t_{s_2}$. From 
(\ref{eq:scaledF2}), we see that if $\nu(1+\kappa_2)(1+\sigma_2)\ll 1$, then $c_2$ 
will be a slow variable with respect to the $t_{c_1}$ timescale. In fact, it is worth 
pointing out that
\begin{equation}
\left[ \nu(1+\kappa_2)(1+\sigma_2)\right]^{-1} = \cfrac{e_2^0}{K_{M_2}+s_2^{\max}} \equiv \epsilon,
\end{equation}
which is the analogue of $\varepsilon_1$ for the secondary reaction. Thus, the scaling 
analysis indicates that $c_2$ will be a slow variable over $t_{s_2}$ if $\epsilon \gg 1$, 
which suggests $e_2^0$ should be large in comparison to $K_{M_2}+s_2^{\max}$.

Next, we see from (\ref{eq:scaledE10}) that 
\begin{equation}
\cfrac{\varepsilon_1 \mu_1^{-1}}{(1+\kappa_1)(1+\sigma_1)} = \cfrac{t_{c_1}}{t_{s_2}},
\end{equation}
and thus $c_1$ will be in QSS on the $t_{s_2}$ timescale as long as $t_{c_1}\ll t_{s_2}$. 

From equation (\ref{eq:scaledE00}), it is clear that if $\mu_1(1+\kappa_1)(1+\sigma_1)\ll 1$, 
then $s_1$ will be a slow variable over the $t_{s_2}$ timescale. However, this condition 
is sufficient but not necessary; since $c_1$ is in QSS, we have:
\begin{equation}\label{eq:MMMM}
\dot{s}_1 \simeq -\cfrac{V_1}{K_{M_1}+s_1}s_1.
\end{equation}
If we then rescale (\ref{eq:MMMM}) with respect to $\bar{T}$, we obtain:
\begin{equation}\label{eq:scaledHs1}
\cfrac{d\hat{s}_1}{d\bar{T}} \simeq  -\mu_1 \cfrac{\hat{s_1}(1+\sigma_1)}{1+\sigma_1\hat{s}_1} \geq -\mu_1.
\end{equation}
Thus, given (\ref{eq:scaledHs1}), we see that $\mu_1 \ll 1$ is both necessary \textit{and} 
sufficient for $s_1$ to be a slow variable with respect $t_{s_2}$ when $c_1$ is in QSS. 
Assuming this condition is met, and the RSA holds, we obtain
\begin{subequations}
\begin{align}
s_1 &\simeq s_1^0,\\
c_1 &\simeq c_1^{\max},\\
s_2 &\simeq s_2^{\lambda}\left[1-\exp(-t/t_{s_2})\right],\\
c_2 &\simeq 0,
\end{align}
\end{subequations}
for $t_{c_1} \lesssim t \lesssim t_{s_2}$.

The remaining dimensionless timescale necessary for the completion of the scaling analysis 
is $\bar{\tau}=t/t_{c_2}$. Rescaling yields
\begin{subequations}
\begin{align}
\cfrac{d\hat{s}_1}{d\bar{\tau}}&=\mu_2(1+\kappa_1)(1+\sigma_1)\left [-\hat{s}_1 + (1-\beta_1)\hat{c}_1\hat{s}_1 + \beta_1\alpha_1\hat{c}_1\right],\label{eq:scaledE00a}\\
\varpi\cfrac{d\hat{c}_1}{d\bar{\tau}}&= \hat{s}_1 - (1-\beta_1)\hat{c}_1\hat{s}_1 - \beta_1\hat{c}_1\label{eq:scaledE10b},\\
\nu \cfrac{d \hat{s}_2}{d\bar{\tau}} &=-\hat{s}_2 + (1-\beta_2)\hat{c}_2\hat{s}_2 + \beta_2\alpha_2\hat{c}_2 + r_S \mu_1 \hat{c}_1,\label{eq:scaledE2c}\\
\cfrac{d \hat{c}_2}{d\bar{\tau}} &=(1+\kappa_2)(1+\sigma_2)\left[\hat{s}_2 - (1-\beta_2)\hat{c}_2\hat{s}_2-\beta_2 \hat{c}_2\right]\label{eq:scaledF2d},
\end{align}
\end{subequations}
where $\varpi = t_{c_1}/t_{c_2}$. Again, if $\varpi\ll 1$, then $c_1$ is in QSS, in which case
\begin{equation}\label{eq:scaledHs1a}
\cfrac{d\hat{s}_1}{d\bar{\tau}} \simeq  -\mu_2 \cfrac{\hat{s_1}(1+\sigma_1)}{1+\sigma_1\hat{s}_1} \geq -\mu_2,
\end{equation}
and $s_1$ is a slow variable with respect to $t_{c_2}$. 

\begin{figure}[htb!]
\centering
\includegraphics[width=12cm]{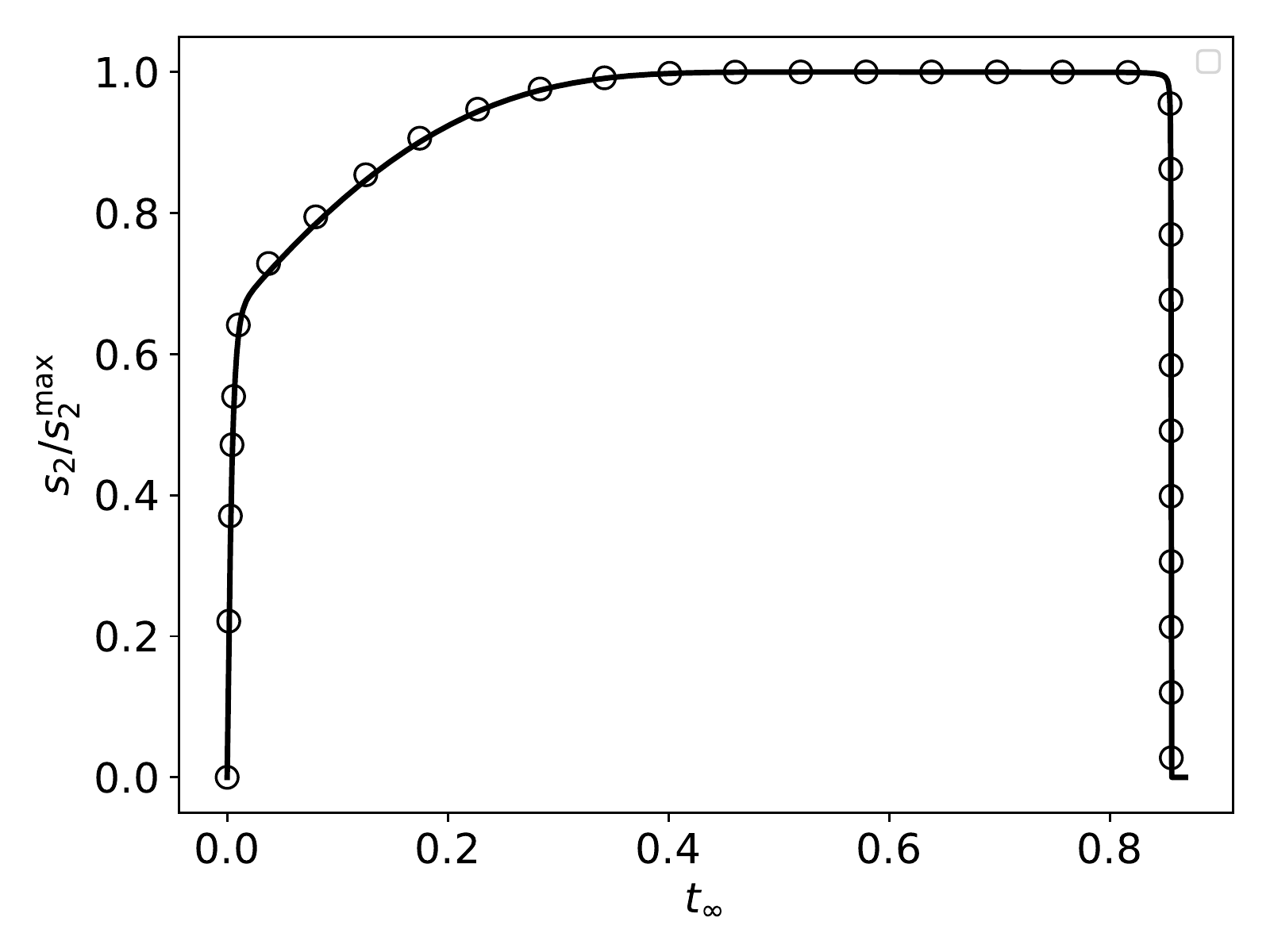}
\caption{\textbf{A graphical illustration of the accuracy of the composite solutions for the 
auxiliary reaction mechanism (\ref{eq:react2})--(\ref{eq:react4})}. The solid black curve
is the numerical solution to (\ref{eq:MA3}), and the unfilled circles mark the composite 
solution~(\ref{eq:compS2}). The constants (without units) used in the numerical simulation 
are: $e_1^0=1$, $s_1^0=1000$, $e_2^0=100$, $k_1=1$, $k_2=1$, $k_3=1$, $k_{-3}=1$, $k_4=2$ 
and $k_{-1}=1$. Time has been mapped to the $t_{\infty}$ scale: 
$t_{\infty}(t) = 1-1/\ln[t+\exp(1)]$. The substrate concentration has been scaled by its 
maximum value.}
\label{fig:8}
\end{figure}

Next, if $\nu \ll 1$, then $s_2$ is in QSS on the $t_{c_2}$ timescale, which implies
\begin{equation}
s_2 \simeq \cfrac{k_{-3}c_2 + k_2c_1^{\max}}{k_3(e_2^0-c_2)}, \;\; \therefore \;\; s_2^{\lambda} \leq s_2 \leq s_2^{\max}.
\end{equation}
Thus, the scaling analysis indicates that
\begin{subequations}
\begin{align}
s_1 &\simeq s_1^0,\\
c_1 &\simeq c_1^{\max},\\
s_2 &\simeq \cfrac{k_{-3}c_2 + k_2c_1^{\max}}{k_3(e_2^0-c_2)},\\
c_2 &\simeq c_2^{\max}\left[1-\exp(-t/t_{c_2})\right],
\end{align}
\end{subequations}
for $t_{s_2} \lesssim t \lesssim t_{c_2}$, and the results of the complete scaling analysis 
allow us to formally construct the composite solutions 
for $s_2$ and $c_2$:
\begin{subequations}
\begin{align}
s_2^{io} &= -s_2^{\lambda}\left[\exp(\displaystyle -t/t_{s_2})\right] + \cfrac{k_{-3}c_2^{io}+k_2c_1^{\max}}{k_3(e_2^0-c_2^{io})}-\cfrac{K_{M_2}}{V_2+\dot{s}_1^{\varepsilon}}\dot{s}_1^{\varepsilon} -s_2^{\max}\label{eq:compS2},\\
c_2^{io} &=-c_2^{\max}\left[\exp(-t/t_{c_2})\right]-\dot{s}_1^{\varepsilon}/k_4\label{eq:compC2}.
\end{align}
\end{subequations}
Together, (\ref{eq:compS2}) and (\ref{eq:compC2}) provide a uniform asymptotic expansion 
that is valid for all time (see {\sc Figure}~\ref{fig:8}).

\subsection{The lag time appears when there are multiple layers and multiple matching timescales}
\begin{figure}[hbt!]
\centering
\includegraphics[width=12cm]{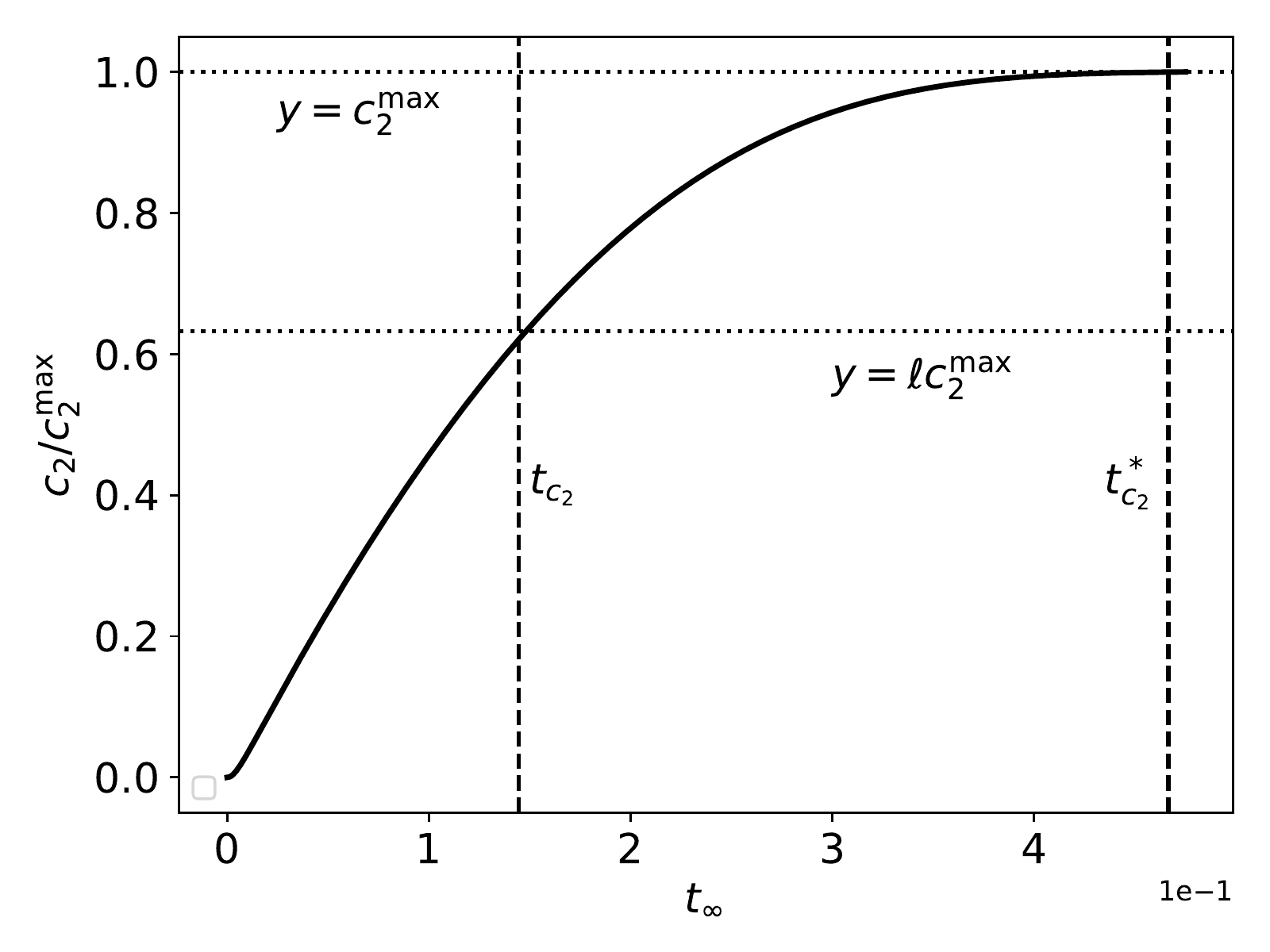}\label{fig:sub7A}
\caption{\textbf{The timescale $t_{c_2}$ is characteristic of the time it takes $c_2$ 
to reach $c_2^{\max}$, and the timescale $t_{c_2}^{\ast}$ is the approximate time it 
takes $c_2$ to reach $c_2^{\max}$, respectively, in the auxiliary reaction 
mechanism~(\ref{eq:react2})--(\ref{eq:react4})}. The thick black curve is the 
numerically-integrated solutions to the mass action 
equations~(\ref{eq:MA1})--(\ref{eq:MA4}). The leftmost dashed vertical line corresponds 
to $t_{c_2}$, and the rightmost dashed vertical line corresponds to 
$t_{c_2}^{\ast} =-t_{c_2}\ln t_{c_2}/t_{s_1}$. The lower dotted horizontal line 
corresponds to the scaled characteristic value $\ell c_2^{\max}$, and the upper 
dotted horizontal line corresponds to $c_2^{\max}$. The constants (without units) 
used in the numerical simulation are: $e_1^0=1$, $s_1^0=1000$, $e_2^0=100$, $k_1=1$, 
$k_2=1$, $k_3=1$, $k_{-3}=1$, $k_4=2$ and $k_{-1}=1$. Time has been mapped to the 
$t_{\infty}$ scale: $t_{\infty}(t) = 1-1/\ln[t+\exp(1)]$, and $c_2$ has been numerically 
scaled by its maximum value. Note that the mass action equations have only been 
integrated from $t=0$ to $t\approx t_{c_2}^{\ast}$ for clarity.}
\label{fig:9}
\end{figure}

In the previous subsection, we derived inner (initial fast transient) and outer 
(quasi-steady-state phase) solutions that are valid when 
$t_{c_1}\ll t_{s_2}\ll t_{c_2} \ll t_{s_1}$. Formally, the ordering, 
$t_{c_1}\ll t_{s_2}\ll t_{c_2} \ll t_{s_1}$, categorizes $t_{c_1}$ as a 
\textit{super-fast} timescale, $t_{s_2}$ as a \textit{fast} timescale, $t_{c_2}$ 
as a \textit{slow} timescale, and $t_{s_1}$ as a \textit{super-slow} timescale. 
From a theoretical perspective, there is utility in estimating the time it takes 
for $s_2$ and $c_2$ to reach $\boldsymbol{x}^{\ast}$, at which time the rate of 
product formation, $\dot{p}$, is at its maximum value. Let $t_{s_2}^{\ast}$ denote 
the actual time it takes $s_2$ to reach $s_2^{\ast}$, and let $t_{c_2}^{\ast}$ denote 
the actual time it takes $s_2$ and $c_2$ to reach $\boldsymbol{x}^{\ast}$. Since 
$t_{s_2}$ and $t_{c_2}$ are characteristic timescales, utilizing them as 
\textit{matching} timescales is problematic since the transition regimes, 
$t_{s_2} \leq t \leq t_{s_2}^{\ast}$ and $t_{c_2} \leq t \leq t_{c_2}^{\ast}$, 
can be quite large. Thus, what we seek are reliable estimates for $t_{s_2}^{\ast}$ 
and $t_{c_2}^{\ast}$. To construct these estimates, we will utilize the approximation 
techniques introduced in Section $3$. Starting with $t_{c_2}^{\ast}$, the inner solution 
for the formation of $c_2$ is
\begin{equation}\label{eq:c2In}
c_2 \simeq c_2^{\max}\left[1-\exp(-t/t_{c_2})\right].
\end{equation}
Although $t_{c_2}$ is a slow timescale, it is fast with respect to $t_{s_1}$. 
Thus, rewriting (\ref{eq:c2In}) with respect to $T$ yields
\begin{equation}
c_2 \simeq c_2^{\max}\left[1-\exp(-T/\mu_2)\right],
\end{equation}
and we see that $T = \mu_2 |\ln \mu_2|$ provides an estimate for $t_{c_2}^{\ast}$:
\begin{equation}\label{eq:LAG}
t_{c_2}^{\ast} \approx -t_{c_2}\ln \mu_2.
\end{equation}
The estimate given in (\ref{eq:LAG}) is the approximate time it takes for 
$\dot{p}$ to reach it maximum with respect to the timescale ordering
$t_{c_1}\ll t_{s_2} \ll t_{c_2} \ll t_{s_1}$ (see {\sc Figure}~\ref{fig:9}).
Formally, the matching timescale $t_{c_2}^{\ast}$ is the lag time, or the time during which 
the second reaction ``lags" behind the first reaction. 

Next we estimate the matching timescale $t_{s_2}^{\ast}$, which is roughly the time it takes for $s_2$ to reach 
QSS. The inner solution
\begin{equation}
s_2 \simeq s_2^{\lambda}\left[1-\exp(-t/t_{s_2})\right],
\end{equation}
can be expressed in terms of its corresponding slow timescale $\bar{\tau}$:
\begin{equation}
s_2 \simeq s_2^{\lambda}\left[1-\exp(-\bar{\tau}/\nu)\right].
\end{equation}
Employing a direct method yields
\begin{equation}\label{eq:direct}
t_{s_2}^{\ast} \approx -t_{s_2}\ln \nu,
\end{equation}
which we take as our approximation to the matching timescale $t_{s_2}^{\ast}.$

In addition to the estimate (\ref{eq:direct}) obtained by the direct method, we can 
employ scaling and justify both (\ref{eq:direct}) and (\ref{eq:LAG}) by invoking 
Theorem~\ref{eq:theorem1}. The scaled mass action equations, 
(\ref{eq:scaled0})--(\ref{eq:scaledB}), can be systematically reduced on the super-slow 
timescale (i.e., $T=t/t_{s_1}$). Since $\varepsilon_2$ and $\mu_1$ are, respectively, 
the smallest parameters with respect to the ordering 
$t_{c_1}\ll t_{s_2}\ll t_{c_2} \ll t_{s_1}$, we can write
\begin{subequations}
\begin{align}
\cfrac{d\hat{s}_1}{dT} &= -\cfrac{\hat{s}_1(\sigma_1+1)}{1+\sigma_1\hat{s_1}}+\mathcal{O}(\varepsilon_2)\label{eq:RA},\\
\mu_2\cfrac{d \hat{c}_2}{dT} &= \cfrac{\hat{s}_1(\sigma_1+1)}{1+\sigma_1\hat{s_1}}-\hat{c}_2 + \mathcal{O}(\varepsilon_2,\mu_1),\label{eq:RB}
\end{align}
\end{subequations}
which are the scaled, leading-order asymptotic equations on the $T$-timescale. Applying 
Theorem~\ref{eq:theorem1} to (\ref{eq:RA})--(\ref{eq:RB}) suggests that $c_2$ should 
approximately reach QSS when $T\approx \mu_2|\ln \mu_2|$.

\begin{figure}[hbt!]
\centering
\includegraphics[width=12cm]{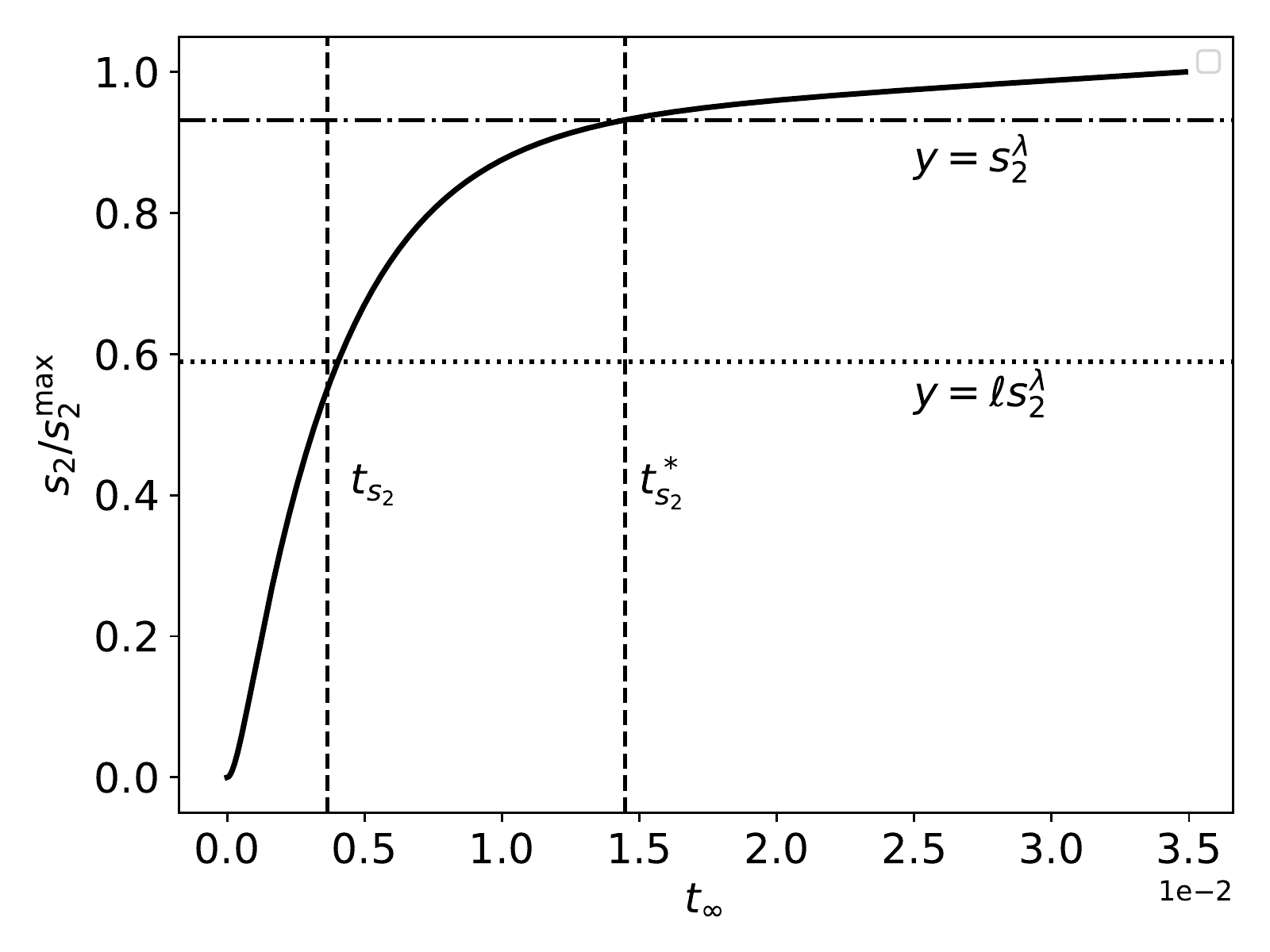}
\caption{\textbf{The timescale $t_{s_2}$ is characteristic of the time it takes $s_2$ 
to reach $s_2^{\lambda}$, and the timescale $t_{s_2}^{\ast}$ is the approximate time it 
takes $s_2$ to reach $s_2^{\lambda}$, respectively, in the auxiliary reaction 
mechanism~(\ref{eq:react2})--(\ref{eq:react4})}. The thick black curve is the 
numerically-integrated solution to the mass action equations~(\ref{eq:MA1})--(\ref{eq:MA4}). 
The leftmost dashed vertical line corresponds to $t_{s_2}$, and the rightmost dashed 
vertical line corresponds to $t_{s_2}^{\ast} =-t_{s_2}\ln t_{s_2}/t_{c_2}$. The lower 
dotted horizontal line corresponds to the scaled characteristic value 
$\ell s_2^{\lambda}$, and the upper dashed/dotted vertical line corresponds to 
$s_2^{\lambda}$. The constants (without units) used in the numerical simulation are: 
$e_1^0=1$, $s_1^0=1000$, $e_2^0=100$, $k_1=1$, $k_2=1$, $k_3=1$, $k_{-3}=1$, 
$k_4=2$ and $k_{-1}=1$. Time has been mapped to the $t_{\infty}$ scale: 
$t_{\infty}(t) = 1-1/\ln[t+\exp(1)]$, and $s_2$ has been numerically-scaled by its 
maximum value. For clarity, the mass action equations have been integrated from 
$t=0$ to $t \approx t_{s_2}^*$.}
\label{fig:10}
\end{figure}

Alternatively, by looking carefully at the scaling obtained with respect to $\bar{\tau}$, 
the leading order dynamics are given by:
\begin{subequations}
\begin{align}
\nu \cfrac{d \hat{s}_2}{d\bar{\tau}} &=\left[-\hat{s}_2 + (1-\beta_2)\hat{c}_2\hat{s}_2 + \beta_2\alpha_2\hat{c}_2\right] + r_S \mu_1\left[1+\mathcal{O}(\mu_2,\varepsilon_1)\right],\label{eq:scaledd3}\\
\cfrac{d \hat{c}_2}{d\bar{\tau}} &=(1+\kappa_2)(1+\sigma_2)\left[\hat{s}_2 - (1-\beta_2)\hat{c}_2\hat{s}_2-\beta_2 \hat{c}_2\right]\label{eq:scaledd4}.
\end{align}
\end{subequations}
Pursuant to Theorem \ref{eq:theorem1}, (\ref{eq:scaledd3})--(\ref{eq:scaledd4}) indicate 
$s_2$ should reach QSS when $\bar{\tau} \approx \nu |\ln \nu|$; consequently, we take
\begin{equation}
t_{s_2}^{\ast} \approx -t_{s_2}\ln \nu,
\end{equation}
as the asymptotic estimate (i.e., the matching timescale) of the time it takes for $s_2$ to reach QSS (see 
{\sc Figure}~\ref{fig:10}).

\subsection{Scaling Analysis: $t_{s_2},t_{c_2} \ll t_{c_1}\ll t_{s_1}$}
\begin{figure}[hbt!]
\centering
\includegraphics[width=12cm]{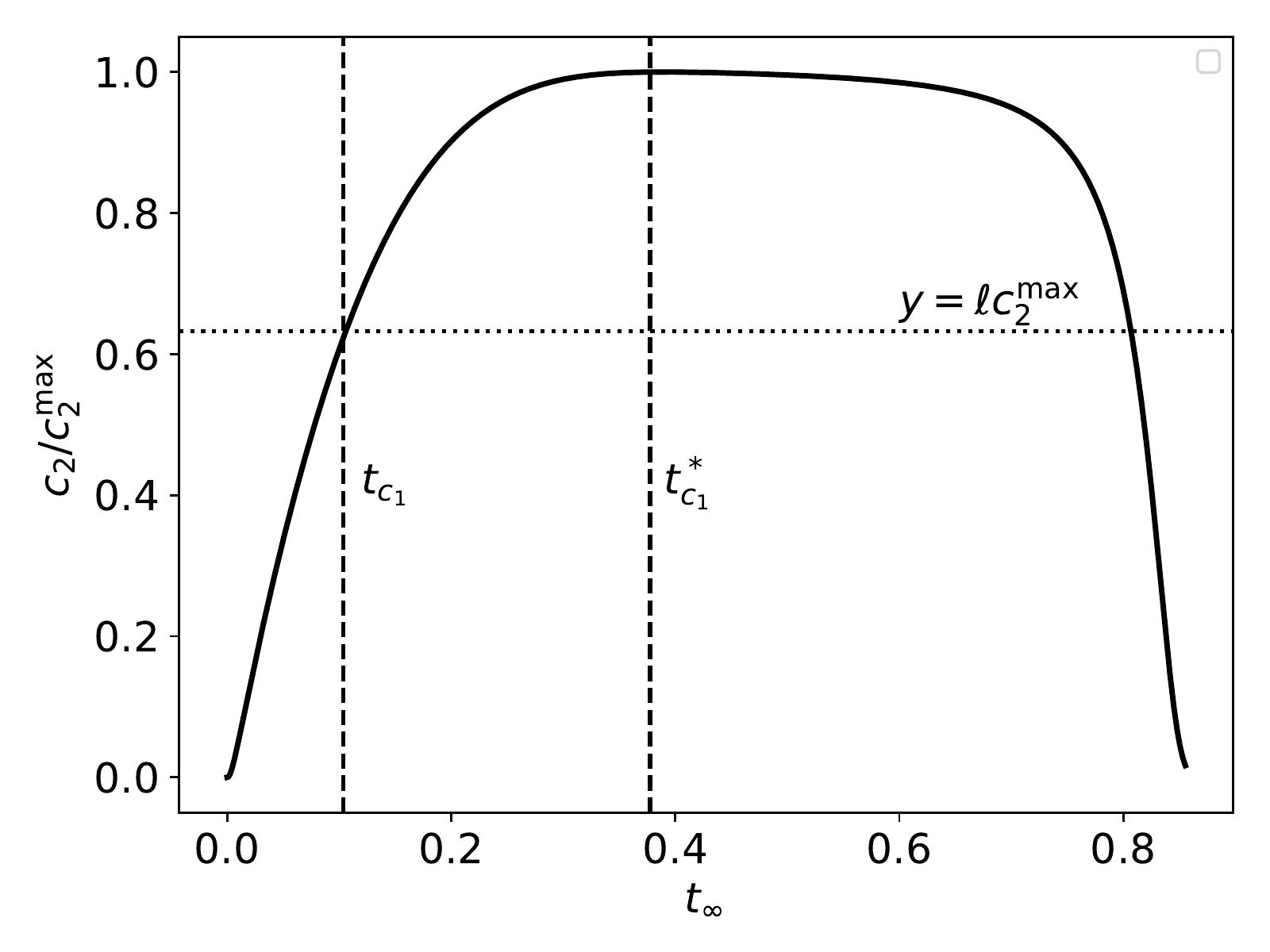}
\caption{\textbf{When $t_{c_2},t_{s_2}\ll t_{c_1}$, the timescale $t_{c_1}$ is characteristic 
of the time it takes $\dot{p}$ to reach its maximum,  and the timescale $t_{c_1}^{\ast}$ is 
the approximate time it takes $c_2$ to reach its maximum, respectively, in the auxiliary reaction 
mechanism~(\ref{eq:react2})--(\ref{eq:react4})}. The thick black curve is the numerically-integrated 
solution to the mass action equations~(\ref{eq:MA1})--(\ref{eq:MA4}). The leftmost dashed vertical 
line corresponds to $t_{c_1}$, and the rightmost dashed vertical line corresponds to 
$t_{c_1}^{\ast} =-t_{c_1}\ln t_{c_1}/t_{s_1}$. The lower dotted horizontal line corresponds 
to the scaled characteristic value $\ell s_2^{\lambda}$. The constants (without units) used 
in the numerical simulation are: $e_1^0=1$, $s_1^0=100$, $e_2^0=100$, $k_1=0.01$, $k_2=1$, 
$k_3=10$, $k_{-3}=1$, $k_4=100$ and $k_{-1}=1$. Time has been mapped to the $t_{\infty}$ 
scale: $t_{\infty}(t) = 1-1/\ln[t+\exp(1)]$, and $c_2$ has been numerically scaled by its 
maximum value.}
\label{fig:11}
\end{figure}

In the most extreme case, when both $t_{s_2}$ and $t_{c_2}$ are much less than $t_{c_1}$ 
in magnitude, scaling analysis indicates that both $s_2$ and $c_2$ are fast variables 
over both the $\tau$ and $T$ timescales:
\begin{subequations}
\begin{align}
\cfrac{d\hat{s}_1}{d\tau}&=\varepsilon_1\left [-\hat{s}_1 + (1-\beta_1)\hat{c}_1\hat{s}_1 + \beta_1\alpha_1\hat{c}_1\right],\label{eq:scaledM1}\\
\cfrac{d\hat{c}_1}{d\tau}&= \left[\hat{s}_1 - (1-\beta_1)\hat{c}_1\hat{s}_1 - \beta_1\hat{c}_1\right],\label{eq:scaledEM2}\\
\lambda_1^{-1}\cfrac{d \hat{s}_2}{d\tau} &=\left[-\hat{s}_2 + (1-\beta_2)\hat{c}_2\hat{s}_2 + \beta_2\alpha_2\hat{c}_2\right] + r_S \mu_1 \hat{c}_1,\label{eq:scaledM3}\\
\lambda_2^{-1}\cfrac{d \hat{c}_2}{d\tau} &=(1+\kappa_2)(1+\sigma_2)\left[\hat{s}_2 - (1-\beta_2)\hat{c}_2\hat{s}_2-\beta_2 \hat{c}_2\right]\label{eq:scaledM4}.
\end{align}
\end{subequations}
Recall that $\lambda_1 \equiv t_{c_1}/t_{s_2}$ and $\lambda_2 \equiv t_{c_1}/t_{c_2}$, 
and that $\lambda_1^{-1}$ and $\lambda_2^{-1}$ will be small when $t_{s_2}$ and $t_{c_2}$ 
are \textit{super-fast} timescales, $t_{c_1}$ is a \textit{fast} timescale, and $t_{s_1}$ 
is a \textit{slow} timescale. Consequently, both $s_2$ and $c_2$ are given in terms of 
$c_1$
\begin{subequations}
\begin{align}
s_2 &\simeq \cfrac{K_{M_2}k_2c_1}{V_2-k_2c_1},\\
c_2 &\simeq \cfrac{k_2{c_1}}{k_4},
\end{align}
\end{subequations}
for $t\geq 0$. Since the secondary reaction is asymptotically determined by $c_1$ when 
$t_{s_2},t_{c_2} \ll t_{c_1}\ll t_{s_1}$, the production rate will reach a maximum when 
$t \approx t_{c_1}^{\ast}$ (see {\sc Figure}~\ref{fig:11}). Thus, the matching timescale $t_{c_1}^*$ is synonymous with time it takes for $\dot{p}$ to reach its maximum value. 

\section{Alternative Orderings of timescale for the auxiliary enzyme reaction}
The previous sections and subsections dealt primarily with the ordering 
$t_{c_1} \ll t_{s_2}\ll t_{c_2} \ll t_{s_1}$. It is natural to ask what happens when 
this ordering starts to change, and in this section we will briefly analyze the dynamics 
of (\ref{eq:react4}) in regimes where the ordering, 
$t_{c_1} \ll t_{s_2}\ll t_{c_2} \ll t_{s_1},$ is no longer preserved.

\subsection{Scaling Analysis for $t_{c_1} \ll t_{c_2}\ll t_{s_2} \ll t_{s_1}$: A three versus four timescale perspective}
\begin{figure}[hbt!]
\centering
\includegraphics[width=12cm]{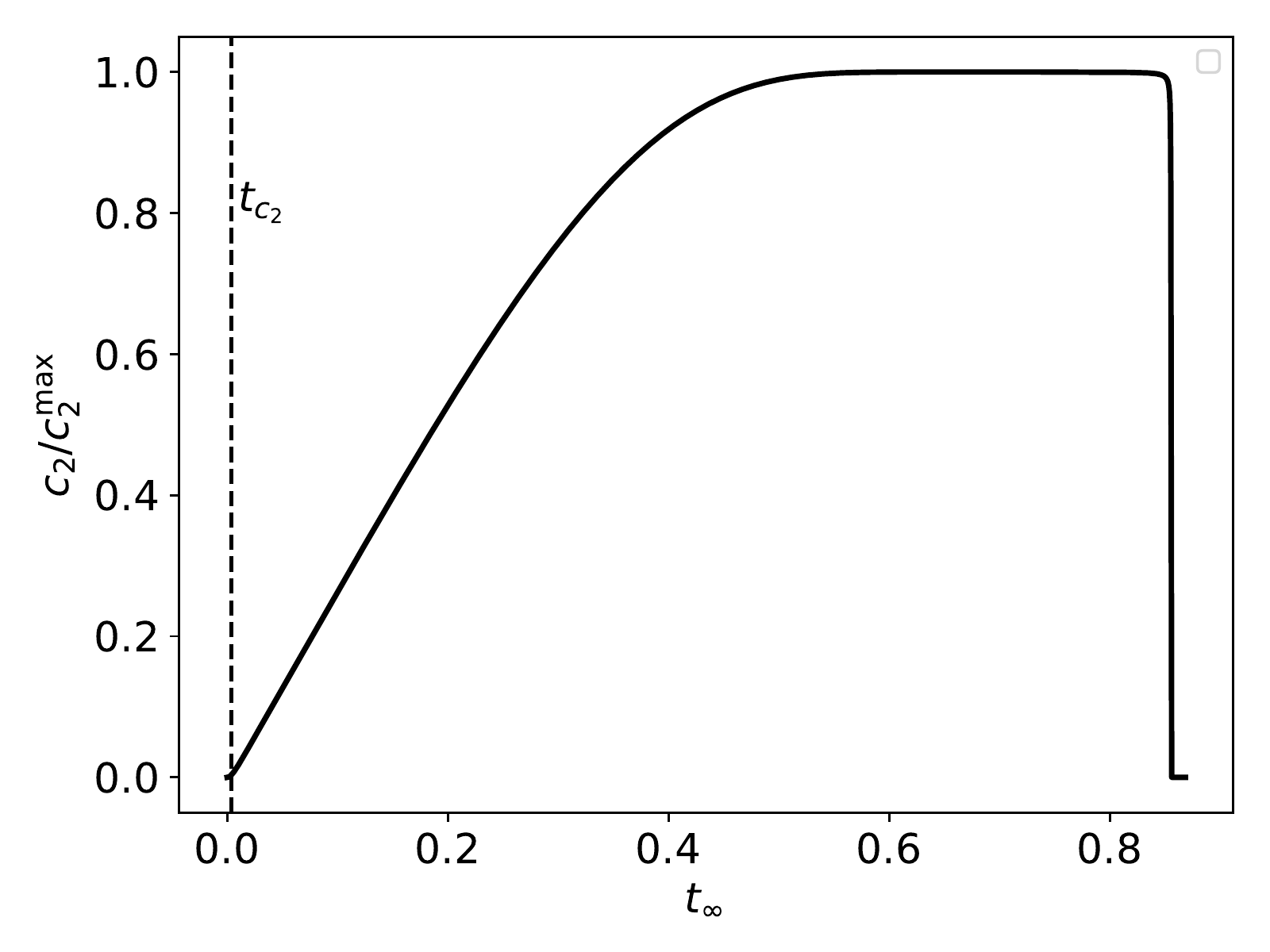}
\caption{\textbf{No significant change in the concentration of $s_2$ or $c_2$ occurs 
over the timescale $t_{c_2}$ in the auxiliary reaction 
mechanism~(\ref{eq:react2})--(\ref{eq:react4}) when 
$t_{c_1} \ll t_{c_2} \ll t_{s_2} \ll t_{s_1}$}. The thick black curve is the 
numerically-integrated solutions to the mass action equations~(\ref{eq:MA1})--(\ref{eq:MA4}). 
The dashed vertical line corresponds to $t_{c_2}$ Note that there is no significant 
increase in the concentration of the intermediate complex over the $t_{c_2}$ timescale. 
The constants (without units) used in the numerical simulation are: $e_1^0=1$, 
$s_1^0=1000$, $e_2^0=1$, $k_1=1$, $k_2=1$, $k_3=1$, $k_{-3}=1$, $k_4=100$ and 
$k_{-1}=1$. Time has been mapped to the $t_{\infty}$ scale: 
$t_{\infty}(t) = 1-1/\ln[t+\exp(1)]$, and $c_2$ has been scaled its maximum value.}
\label{fig:12}
\end{figure}

The first ordering we consider is that in which $t_{c_1}$ is a \textit{super-fast} timescale, 
$t_{c_2}$ is a fast timescale, $t_{s_2}$ is a slow timescale, and $t_{s_1}$ is 
\textit{super-slow} timescale: $t_{c_1} \ll t_{c_2}\ll t_{s_2} \ll t_{s_1}$. We will 
start the analysis by observing the scaling with respect to $\bar{\tau}$:
\begin{subequations}
\begin{align}
\cfrac{d\hat{s}_1}{d\bar{\tau}}&=\mu_2(1+\kappa_1)(1+\sigma_1)\left [-\hat{s}_1 + (1-\beta_1)\hat{c}_1\hat{s}_1 + \beta_1\alpha_1\hat{c}_1\right],\label{eq:sA}\\
\varpi\cfrac{d\hat{c}_1}{d\bar{\tau}}&= \left[\hat{s}_1 - (1-\beta_1)\hat{c}_1\hat{s}_1 - \beta_1\hat{c}_1\right]\label{eq:sB},\\
\cfrac{d \hat{s}_2}{d\bar{\tau}} &=\nu^{-1}\left[-\hat{s}_2 + (1-\beta_2)\hat{c}_2\hat{s}_2 + \beta_2\alpha_2\hat{c}_2\right] + r_S \mu_2 \hat{c}_1,\label{eq:sC}\\
\cfrac{d \hat{c}_2}{d\bar{\tau}} &=(1+\kappa_2)(1+\sigma_2)\left[\hat{s}_2 - (1-\beta_2)\hat{c}_2\hat{s}_2-\beta_2 \hat{c}_2\right]\label{eq:sD}.
\end{align}
\end{subequations}
If $\nu^{-1}\ll 1$, then we immediately see that
\begin{equation}\label{eq:sLW}
\cfrac{d \hat{s}_2}{d\bar{\tau}} \simeq r_S \mu_2 \hat{c}_1 + \mathcal{O}(\nu^{-1}).
\end{equation}
Next, because we have assumed in our ordering that $t_{c_1} \ll t_{c_2}$, 
equation~(\ref{eq:sLW}) can be reduced further by noting that $\hat{c}_1 \simeq 1$:
\begin{equation}\label{eq:sLW1}
\cfrac{d \hat{s}_2}{d\bar{\tau}} \simeq r_S\mu_2.
\end{equation}
If we can then find a bound on $ r_S \mu_2$ by showing that $r_S \mu_2 \leq K$ and 
$K \sim \nu^{-1}$, then it follows that $s_2$ is a \textit{slow} variable 
with respect to $\bar{\tau}$. Expanding $r_S \lambda_2$ yields
\begin{equation}
r_S\mu_2 = \cfrac{s_1^0}{s_2^{\max}}\cfrac{t_{c_2}}{t_{s_1}} = \cfrac{e_2^0}{K_{M_2}}-\varepsilon_1\cfrac{k_2s_1^0}{K_{M_2}k_4} \geq 0,
\end{equation}
which implies 
\begin{equation}
r_S \mu_2  \leq \cfrac{e_2^0}{K_{M_2}} \equiv K \leq \nu^{-1}.
\end{equation}
Thus, based on the scaling analysis, we take $s_2 \simeq 0$ for $t \leq t_{c_2}$. 
The immediate consequence is that $c_2 \simeq 0$ for $t \leq t_{c_2}$, 
since complex cannot form without the presence of substrate. Thus, no significant 
change in the concentration of $s_2$ or $c_2$ occurs for $t \leq t_{c_2}$ when 
$t_{c_1}\ll t_{c_2}\ll t_{s_2} \ll t_{s_1}$ (see {\sc Figure}~\ref{fig:12}).

Next, we scale with respect to the slow timescale, $\bar{T}$:
\begin{subequations}
\begin{align}
\cfrac{d\hat{s}_1}{d\bar{T}}&=\mu_1(1+\kappa_1)(1+\sigma_1)\left [-\hat{s}_1 + (1-\beta_1)\hat{c}_1\hat{s}_1 + \beta_1\alpha_1\hat{c}_1\right],\label{eq:sB1}\\
\varepsilon_1\cfrac{d\hat{c}_1}{d\bar{T}}&= \mu_1(1+\kappa_1)(1+\sigma_1)\left[\hat{s}_1 - (1-\beta_1)\hat{c}_1\hat{s}_1 - \beta_1\hat{c}_1\right]\label{eq:sB2}\\
\cfrac{d \hat{s}_2}{d\bar{T}} &=-\hat{s}_2 + (1-\beta_2)\hat{c}_2\hat{s}_2 + \beta_2\alpha_2\hat{c}_2 + r_S \mu_1 \hat{c}_1,\label{eq:sB3}\\
\epsilon\cfrac{d \hat{c}_2}{d\bar{T}} &=\hat{s}_2 - (1-\beta_2)\hat{c}_2\hat{s}_2-\beta_2 \hat{c}_2\label{eq:sB4}.
\end{align}
\end{subequations}
The term $r_S \mu_1$ is $\mathcal{O}(1)$, and $\hat{c}_2$ can be approximated as being 
in QSS since $\epsilon \ll 1$ when $t_{c_2} \ll t_{s_2}$. Putting these observations 
together yields the dimensional equation
\begin{equation}\label{eq:104}
\dot{s}_2 \simeq -\cfrac{V_2}{K_{M_2}+s_2}s_2 + k_2c_1^{\max}, \quad t \lesssim t_{s_2},
\end{equation}
which admits an exact solution in the form of a Lambert-W function
\begin{equation}\label{eq:Lambert}
s_2 \simeq s_2^{\max}(1+\psi W\left[-\psi^{-1}\exp(-\psi^{-1}-\Theta\cdot t)\right]), \quad t \lesssim t_{s_2},
\end{equation}
where $\psi \equiv V_2/(k_2c_1^{\max})$ and $\Theta \equiv (V_2-k_2c_1^{\max})^2/(V_2 K_{M_2})$. 
From (\ref{eq:Lambert}), we have a new timescale, $t_{s_2}^{\chi}$:
\begin{equation}
t_{s_2}^{\chi}\equiv \cfrac{K_{M_2}+s_2^{\max}}{V_2}.
\end{equation}

\begin{figure}[hbt!]
\centering
\includegraphics[width=12cm]{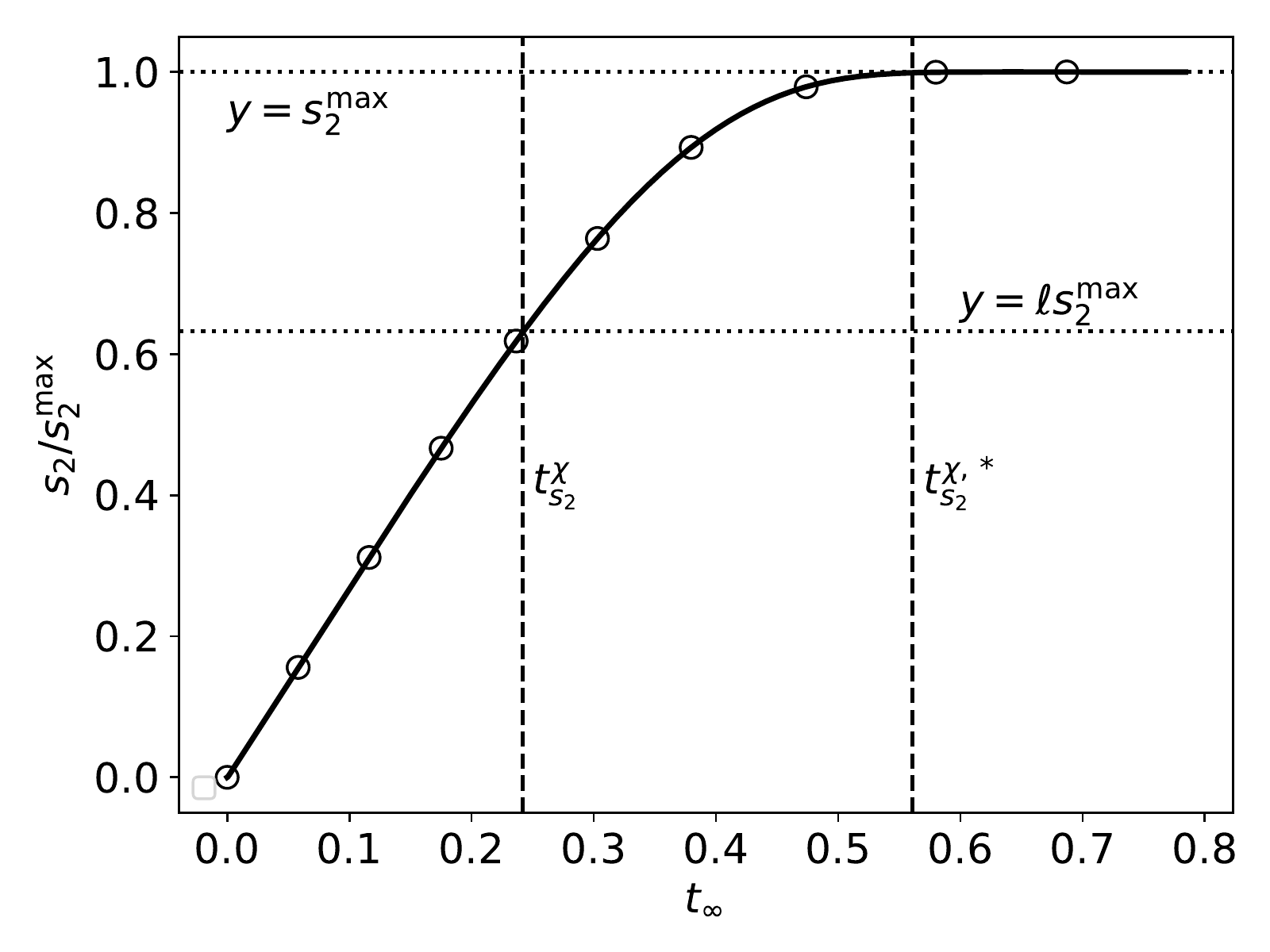}
\caption{\textbf{The validity of $t_{s_2}^{\chi}$ and $t_{s_2}^{\chi,\ast}$ in the auxiliary 
reaction mechanism~(\ref{eq:react2})--(\ref{eq:react4}) when $\epsilon \ll 1$}. The thick 
black curve is the numerically-integrated solution to the mass action 
equations~(\ref{eq:MA1})--(\ref{eq:MA4}), and the unfilled circles mark the inner solution 
given by (\ref{eq:Lambert}). The leftmost dashed vertical line corresponds to $t_{s_2}^{\chi}$, 
and the rightmost dashed vertical line corresponds to 
$t_{s_2}^{\chi,\ast} =-t_{s_2}^{\chi}\ln t_{s_2}^{\chi}/t_{s_1}$.
The lower dotted horizontal line corresponds to $y=\ell s_2^{\max}$, and the upper dotted 
horizontal line corresponds to $y=s_2^{\max}$. The constants (without units) used in the 
numerical simulation are: $e_1^0=1$, $s_1^0=1000$, $e_2^0=1$, $k_1=1$, $k_2=1$, $k_3=1$, 
$k_{-3}=1$, $k_4=100$ and $k_{-1}=1$. Time has been mapped to the $t_{\infty}$ scale: 
$t_{\infty}(t) = 1-1/\ln[t+\exp(1)]$, and $s_2$ has been numerically-scaled by its maximum 
value. Note that the mass action equations have only been integrated from $t=0$ to 
$t\approx t_{s_2}^{\chi,\ast}$ for clarity.}
\label{fig:13}
\end{figure}

Since no significant change in the concentration of any chemical species occurs over 
$t_{c_2}$, the \textit{kinetic} analysis in this regime can be effectively carried out 
with \textit{three} timescales: $t_{c_1},t_{s_2}^{\chi},t_{s_1}$ \citep{Eilertsen:2018:KAS}.
Additionally, it is also worth noting that rescaling the mass action equations with 
respect to $T^{\chi}=t/t_{s_2}^{\chi}$ yields,
\begin{subequations}
\begin{align}
\cfrac{d\hat{s}_1}{dT^{\chi}}&=\cfrac{t_{s_2}^{\chi}}{t_{s_1}}(1+\kappa_1)(1+\sigma_1)\left [-\hat{s}_1 + (1-\beta_1)\hat{c}_1\hat{s}_1 + \beta_1\alpha_1\hat{c}_1\right],\label{eq:sC1}\\
\cfrac{t_{c_1}}{t_{s_2}^{\chi}}\cfrac{d \hat{c}_1}{dT^{\chi}}&= \hat{s}_1 - (1-\beta_1)\hat{c}_1\hat{s}_1 - \beta_1\hat{c}_1\label{eq:sC2},\\
\cfrac{d \hat{s}_2}{dT^{\chi}} &=(1+\kappa_2)(1+\sigma_2)\left[-\hat{s}_2 + (1-\beta_2)\hat{c}_2\hat{s}_2 + \beta_2\alpha_2\hat{c}_2\right] + \hat{c}_1,\label{eq:sC3}\\
\epsilon\cfrac{d \hat{c}_2}{dT^{\chi}} &=(1+\kappa_2)(1+\sigma_2)\left[\hat{s}_2 - (1-\beta_2)\hat{c}_2\hat{s}_2-\beta_2 \hat{c}_2\right]\label{eq:sC4},
\end{align}
\end{subequations}
and the term in front of $\hat{c}_1$ in (\ref{eq:sC3}) is euqal to $1$. It follows 
that (\ref{eq:sC3}) is, to leading order, given by
\begin{equation}
\cfrac{d\hat{s}_2}{dT^{\chi}} \simeq -\cfrac{\hat{s}_2(1+\sigma_2)}{1+\sigma_2\hat{s}_2} + 1.\\
\end{equation}
Furthermore, rescaling the mass action equations with respect to $T$ yields
\begin{subequations}
\begin{align}
\cfrac{d\hat{s}_1}{dT}&=(1+\kappa_1)(1+\sigma_1)\left [-\hat{s}_1 + (1-\beta_1)\hat{c}_1\hat{s}_1 + \beta_1\alpha_1\hat{c}_1\right],\label{eq:sD1}\\
\varepsilon_2\cfrac{d \hat{c}_1}{dT}&= \hat{s}_1 - (1-\beta_1)\hat{c}_1\hat{s}_1 - \beta_1\hat{c}_1\label{eq:sD2},\\
\cfrac{t_{s_2}^{\chi}}{t_{s_1}}\cfrac{d \hat{s}_2}{dT} &=(1+\kappa_2)(1+\sigma_2)\left[-\hat{s}_2 + (1-\beta_2)\hat{c}_2\hat{s}_2 + \beta_2\alpha_2\hat{c}_2\right] + \hat{c}_1,\label{eq:sD3}\\
\mu_2\cfrac{d \hat{c}_2}{dT} &=(1+\kappa_2)(1+\sigma_2)\left[\hat{s}_2 - (1-\beta_2)\hat{c}_2\hat{s}_2-\beta_2 \hat{c}_2\right]\label{eq:sD4},
\end{align}
\end{subequations}
from which it directly follows (see {\sc Figure}~\ref{fig:13}) that the time it takes 
for $\dot{p}$ to reach its maximum is given by
\begin{equation}\label{eq:SCALEt}
t_{s_2}^{\chi,\ast} \approx -t_{s_2}^{\chi}\ln \cfrac{t_{s_2}^{\chi}}{t_{s_1}}.
\end{equation}
The timescale (\ref{eq:SCALEt}) is the matching timescale for $s_2$. It is a very good estimate of the time it takes $s_2$ to reach QSS, and corresponds to the time it takes the phase-plane trajectory to reach $\boldsymbol{x}^*$ when $\epsilon \ll 1$ \citep{Eilertsen:2018:KAS}.

\subsection{Scaling Analysis: $t_{c_2} \ll t_{c_1}\ll t_{s_2} \ll t_{s_1}$}
In the previous subsection we showed that $t_{c_2}$ was a ``hidden" timescale: no 
significant accumulation of $s_2$ and $c_2$ occurs over $t_{c_2}$ when 
$t_{c_1}\ll t_{c_2} \ll t_{s_2} \ll t_{s_1}$. In this subsection we examine was what 
happens when $t_{c_2} \ll t_{c_1}$. First, note that
\begin{equation}\label{eq:T1}
    \displaystyle \lim_{k_4 \to \infty} t_{s_2}^{\chi} = t_{s_2},
\end{equation}
and second,
\begin{equation}\label{eq:T2}
    \displaystyle \lim_{k_4 \to \infty} c_2^{\max} = 0, \;\;\text{and}\;\;\displaystyle \lim_{k_4 \to \infty} s_2^{\max} = s_2^{\lambda}.
\end{equation}
Finally, since
\begin{equation}\label{eq:T3}
  W\left[-\psi^{-1}\exp(-\psi^{-1}-\Theta\cdot t)\right] \simeq   -\psi^{-1}\exp(-\psi^{-1}-\Theta\cdot t),\quad \psi^{-1} \ll 1,
\end{equation}
we can combine (\ref{eq:T1}), (\ref{eq:T2}) and (\ref{eq:T3}) to yield
\begin{equation}\label{eq:ELM}
    s_2 \approx s_2^{\lambda}\left[1-\exp(-t/t_{s_2})\right], \quad\;\;\text{for}\;\; t \lesssim t_{s_2}.
\end{equation}
From a geometrical point of view, the $c_2$-nullcline gets pressed against the $s_2$-axis 
in the phase-plane as $k_4 \to \infty$, and $c_2^{\max}$ is almost negligible in magnitude. 
Thus, when $k_4 \gg k_3e_2^0$, the mass action kinetics can \textit{essentially} be 
approximated by (\ref{eq:ELM}), since $t_{s_2} \approx t_{s_2}^{\chi}$ and 
$s_2^{\max} \approx s_2^{\lambda}$ as $t_{c_2} \to 0$. Consequently, $t_{s_2}$ is 
approximately characteristic of the time it takes $s_2$ to reach $s_2^{\max}$ in regimes 
where $t_{c_2}$ is a \textit{super-fast} timescale and 
$t_{c_2}\ll t_{c_1} \ll t_{s_2}^{\chi} \ll t_{s_1}$.

\subsection{Scaling Analysis: $t_{s_2} \ll t_{c_1}\ll t_{c_2} \ll t_{s_1}$}
\begin{figure}[hbt!]
\centering
\includegraphics[width=12cm]{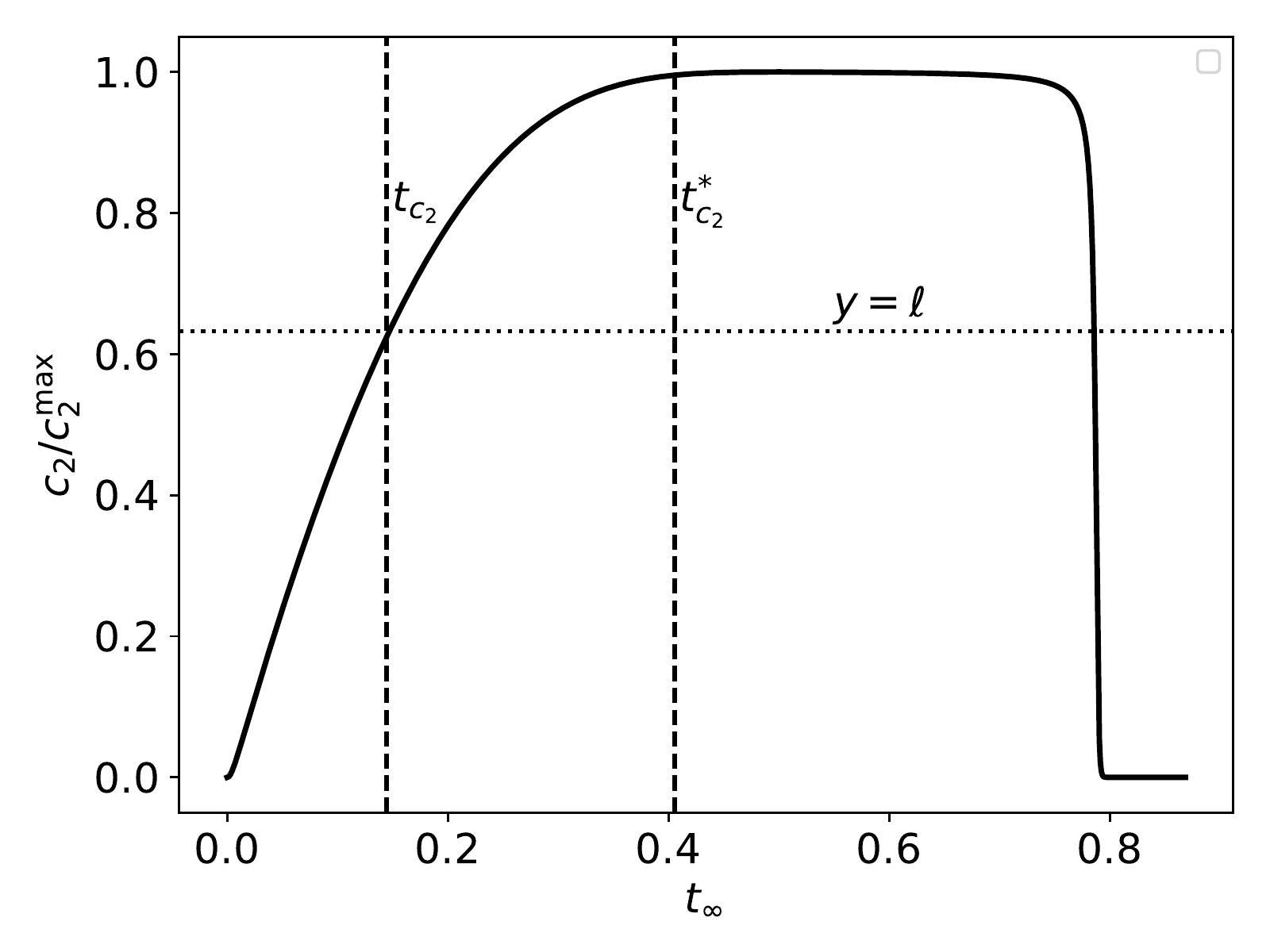}
\caption{\textbf{The lag time in the auxiliary reaction 
mechanism~(\ref{eq:react2})--(\ref{eq:react4}) when 
$t_{s_2} \ll t_{c_1} \ll t_{c_2} \ll t_{s_1}$}. The thick black curve is the 
numerically-integrated solution to the mass action equations (\ref{eq:MA1})--(\ref{eq:MA4}), 
and the unfilled circles mark the inner solution given by (\ref{eq:Lambert}). The leftmost 
dashed vertical line corresponds to $t_{c_2}$, and the rightmost dashed vertical line 
corresponds to $t_{c_2}^{\ast} =-t_{c_2}\ln t_{c_2}/t_{s_1}$. The lower dotted horizontal 
line corresponds to $y=\ell$; The constants (without units) used in the 
numerical simulation are: $e_1^0=1$, $s_1^0=1000$, $e_2^0=1$, $k_1=1$, $k_2=1$, $k_3=1$, 
$k_{-3}=1$, $k_4=100$ and $k_{-1}=1$. Time has been mapped to the $t_{\infty}$ scale: 
$t_{\infty}(t) = 1-1/\ln[t+\exp(1)]$, and $c_2$ has been numerically-scaled by its maximum 
value.}
\label{fig:14}
\end{figure}

Another case is when $t_{s_2}$ is a super-fast timescale. Under this scenerio, the scaled 
equations indicate that $s_2$ is in QSS for the duration of the reaction. Geometrically, 
$s_2$ will closely follow the $s_2$-nullcline as it moves in the $s_2$--$c_2$ phase--plane. 
In this case $c_2$ is asymptotic to
\begin{equation}\label{eq:finalC}
\dot{c}_2 \simeq -k_4c_2 + k_2c_1,\quad t \geq 0,
\end{equation}
and thus $t_{c_2}$ remains characteristic of the time is takes $c_2$ to reach its maximum 
value, and the matching timescale $t_{c_2}^*$ provides an estimate for the time it takes for $\dot{p}$ to reach its maximum value (see {\sc Figure}~\ref{fig:14}).

\subsection{Scaling Analysis: $t_{c_1} \ll t_{c_2}\approx t_{s_2} \ll t_{s_1}$}
\begin{figure}[hbt!]
\centering
\includegraphics[width=12cm]{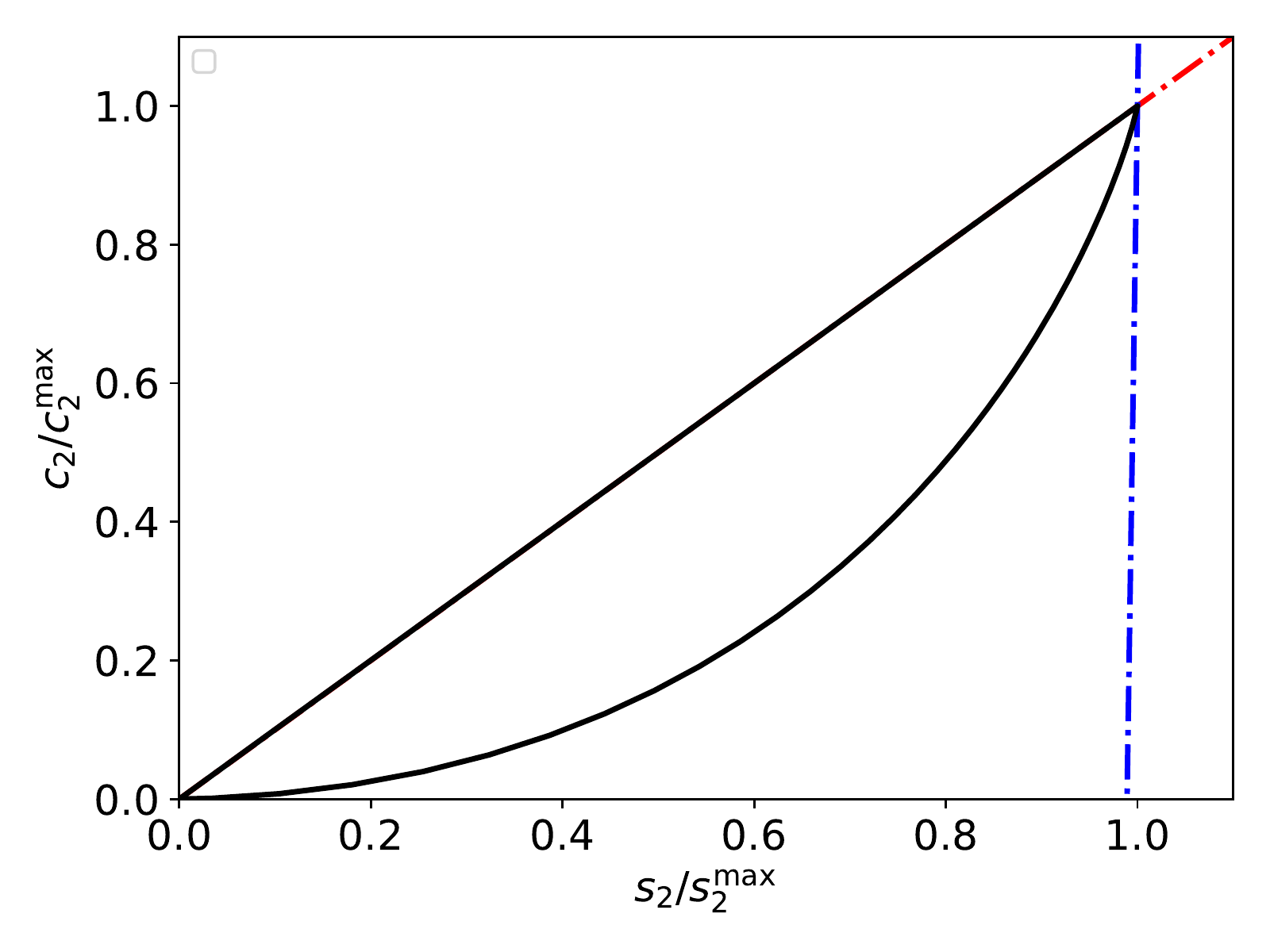}
\caption{\textbf{Phase--plane dynamics of the auxiliary reaction 
mechanism~(\ref{eq:react2})--(\ref{eq:react4}) when 
$t_{c_1} \ll t_{c_2} \approx t_{s_2} \ll t_{s_1}$}. The thick black curve is the 
numerically-integrated solution to the mass action equations~(\ref{eq:MA1})--(\ref{eq:MA4}), 
the dashed/dotted red curve is the $c_2$-nullcline and the dashed/dotted blue curve is 
the stationary $s_2$-nullcline. Notice that the trajectory does not follow a path that 
lies close to either nullcline in the approach to $\boldsymbol{x}^*$. The constants 
(without units) used in the numerical simulation are: $e_1^0=1$, $s_1^0=1000$, $e_2^0=10$, 
$k_1=1$, $k_2=1$, $k_3=10$, $k_{-3}=1$, $k_4=100$ and $k_{-1}=1$.}
\label{fig:15}
\end{figure}

\begin{figure}[hbt!]
\centering
\includegraphics[width=12cm]{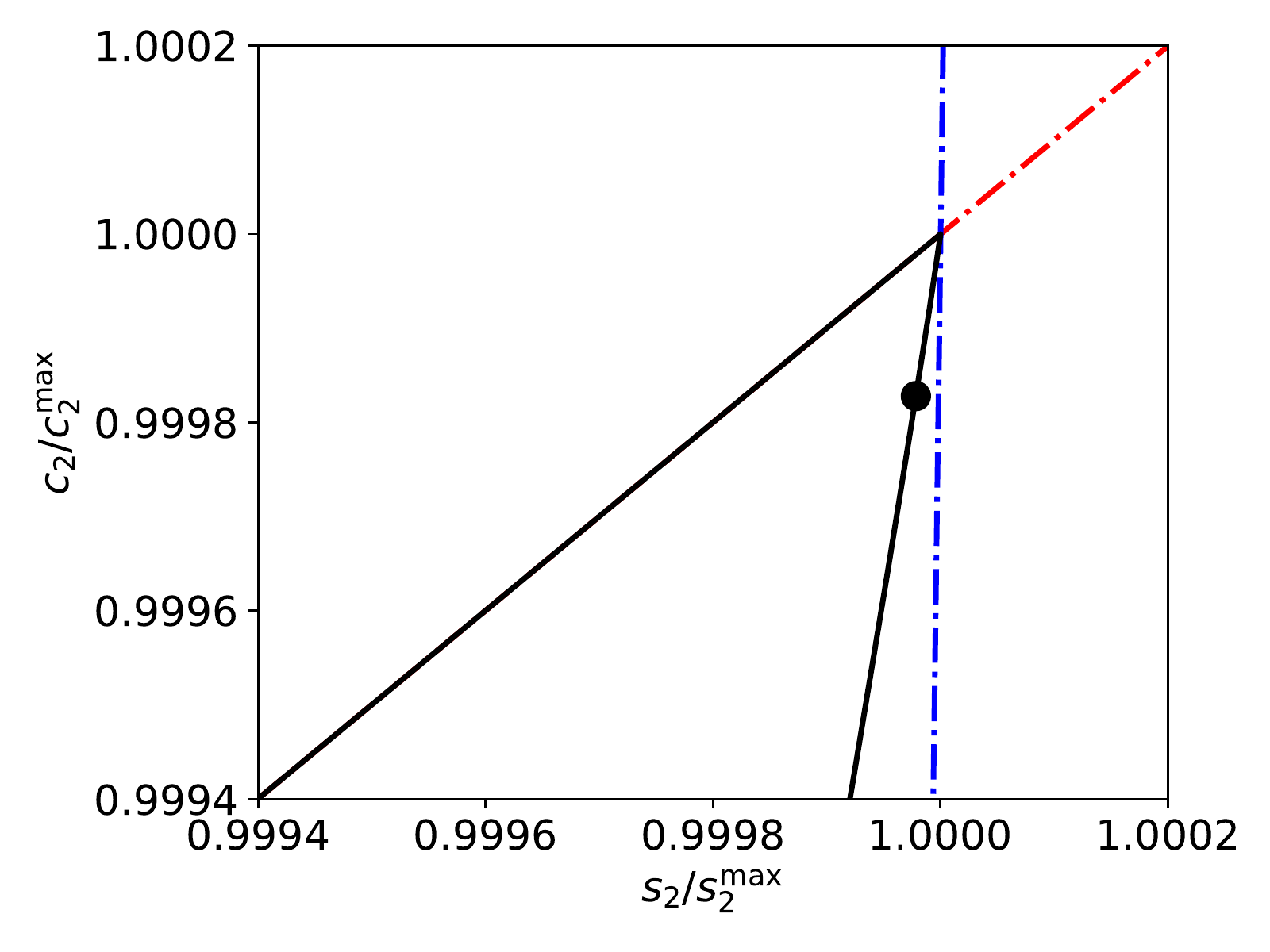}
\caption{\textbf{The lag time in the auxiliary reaction 
mechanism~(\ref{eq:react2})--(\ref{eq:react4}) when $t_{c_1} \ll t_{c_2} = t_{s_2} \ll t_{s_1}$}. 
This is a close-up of {\sc Figure}~\ref{fig:15} near $\boldsymbol{x}^{\ast}$. The thick black 
curve is the numerically-integrated solution to the mass action 
equations~(\ref{eq:MA1})--(\ref{eq:MA4}), the dashed/dotted red curve is the $c_2$-nullcline 
and the dashed/dotted blue curve is the stationary $s_2$-nullcline. The solid black circle 
marks the trajectory when $t=t_{s_2}^{\ast} = -t_{s_2}\ln t_{s_2}/t_{s_1}$. Notice that
Tikhonov's Theorem still provides a reasonable estimate of the lag time, which is synonymous with the matching timescale corresponding to either $s_2$ or $c_2$. The constants 
(without units) used in the numerical simulation are: $e_1^0=1$, $s_1^0=1000$, $e_2^0=10$, 
$k_1=1$, $k_2=1$, $k_3=10$, $k_{-3}=1$, $k_4=100$ and $k_{-1}=1$.}
\label{fig:16}
\end{figure}

Up until this point, we have been able to derive characteristic timescales that quantify 
the temporal order of magnitude of a specific trajectory's rapid approach to QSS. Our 
success in the derivation of characteristic timescales resides in the fact that, so far, 
we have only considered regimes in which trajectories are asymptotic to one-dimensional 
manifolds (i.e., the $s_2$-nullcline or the $c_2$-nullcline) in their approach to the 
zero-dimensional manifold, $\boldsymbol{x}^{\ast}$. However, their are many such 
trajectories that are not asymptotic to a particular manifold in the approach to
$\boldsymbol{x}^{\ast}$. For example, if $t_{c_1} \ll t_{c_2}\approx t_{s_2} \ll t_{s_1}$, 
then it is obvious from both the scaling analysis and the phase-plane dynamics that 
the trajectory will not follow closely to either nullcline in its approach to 
$\boldsymbol{x}^{\ast}$ (see {\sc Figure}~\ref{fig:15}).
It is not obvious in this case how to go about determining the lag time. However, 
Theorem~\ref{eq:theorem1} suggests that either matching timescale $t_{c_2}^*$ or $t_{s_2}^*$ should yield 
a reasonable approximation to the lag time. Thus, even though the transient solution is 
unknown, the scaling analysis still provides a good estimate of the time it takes for 
the secondary reaction to ``catch" the primary reaction and for $\dot{p}$ to reach its maximum value (see {\sc Figure}~\ref{fig:16}).

\section{The region of validity of the timescale estimations}
\begin{figure}[hbt!]
\centering
\includegraphics[width=12cm]{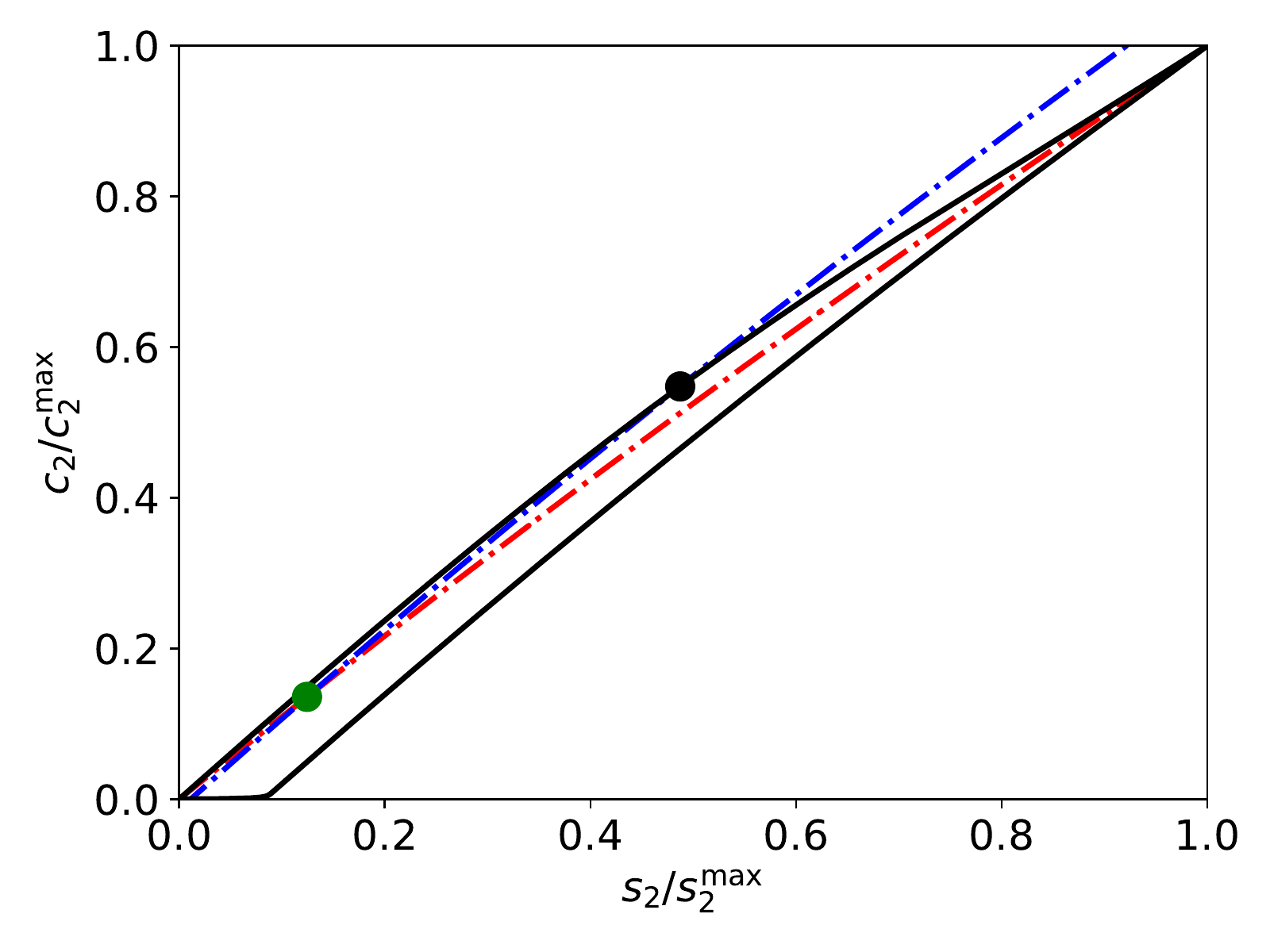}
\caption{\textbf{The trajectory follows the $s_2$-nullcline in the phase--plane of the auxiliary
reaction mechanism~(\ref{eq:react2})--(\ref{eq:react4}) when $t_{s_2}^{\chi}/t_{s_1}\ll 1$}. The 
thick black curve is the numerically-integrated solution to the mass action 
equations~(\ref{eq:MA1})--(\ref{eq:MA4}), the dashed/dotted red curve is the $c_2$-nullcline and 
the dashed/dotted blue curve is a snapshot of $s_2$-nullcline when $t\approx 1.1\cdot t_{s_1}$. 
The black dot is the corresponding snapshot of the numerical solution to
(\ref{eq:MA1})--(\ref{eq:MA4}). In this simulation, 
$t_{s_2}^{\chi}/t_{s_1} \approx 0.001 < \mu_2 \approx 0.1$; consequently, the trajectory 
follows the $s_2$-nullcline but fails to closely follow $\boldsymbol{x}^*$ (see 
{\sc Movie}~1 in Supplementary Materials). The constants (without units) used in the 
numerical simulation are: $e_1^0=1$, $s_1^0=100$, $e_2^0=100$, $k_1=1$, $k_2=1$, $k_3=1$, 
$k_{-3}=1$, $k_4=0.1$ and $k_{-1}=1$.}
\label{fig:17}
\end{figure}

\begin{figure}[hbt!]
\centering
\includegraphics[width=12cm]{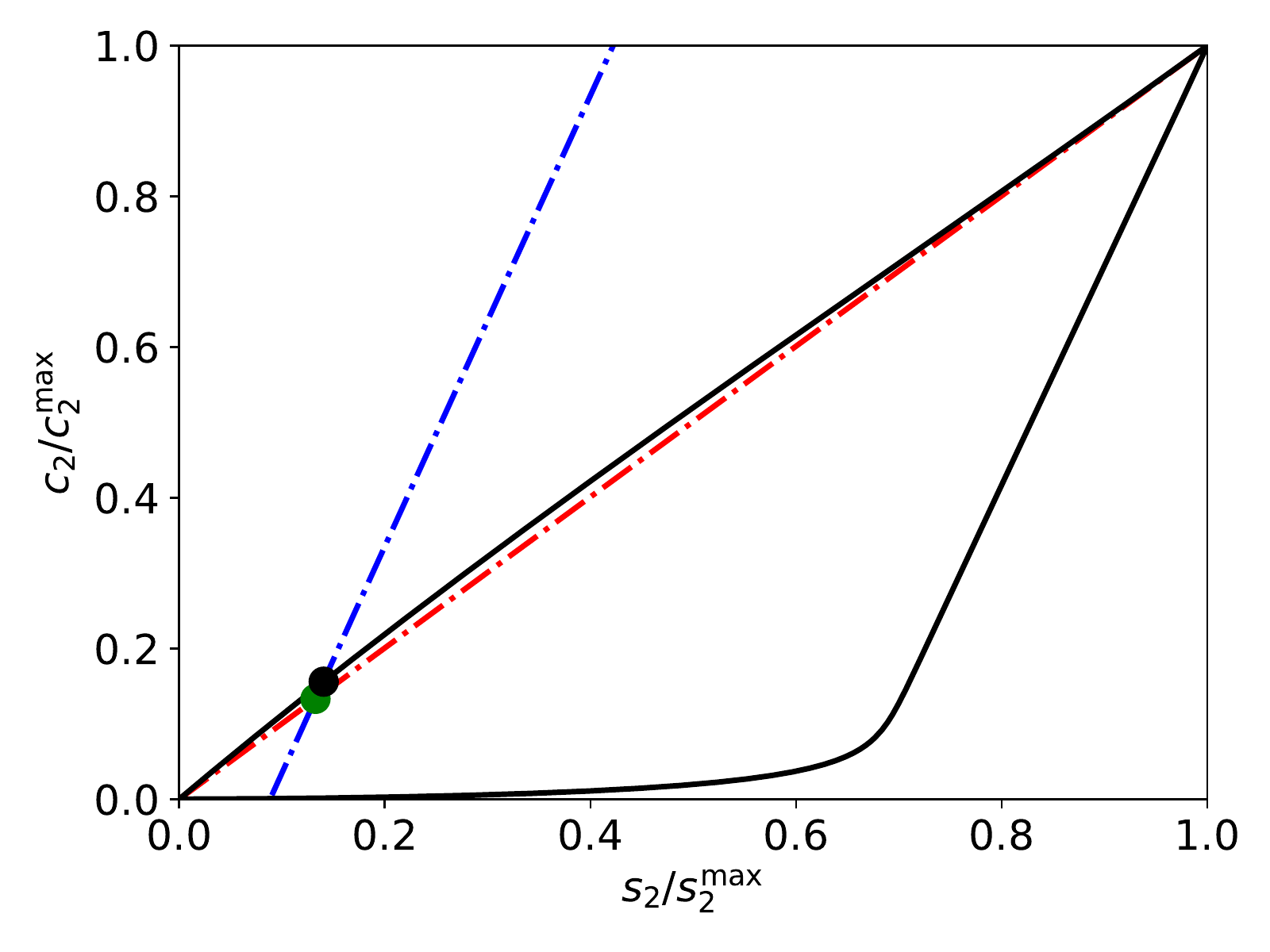}
\caption{\textbf{The component-wise error when the indicator reaction is fast in the auxiliary
reaction mechanism~(\ref{eq:react2})--(\ref{eq:react4})}. The thick black curve is the
numerically-integrated solution to the mass action equations~(\ref{eq:MA1})--(\ref{eq:MA4}), 
the dashed/dotted red curve is the $c_2$-nullcline and the dashed/dotted blue curve is a 
snapshot of $s_2$-nullcline when $t\approx 1.1\cdot t_{s_1}$. The black dot is the 
corresponding snapshot of the numerical solution to (\ref{eq:MA1})--(\ref{eq:MA4}). In this 
simulation, $t_{s_2}^{\chi}/t_{s_1} \approx 0.0001 < \mu_2 \approx 0.005$; consequently, 
the trajectory sits ``just behind" and slightly above $\boldsymbol{x}^*$ (green dot) since 
the trajectory will be closer to the $s_2$-nullcline than the $c_2$-nullcline (see 
{\sc Movie}~2 in Supplementary Materials). The constants (without units) used in the 
numerical simulation are: $e_1^0=1$, $s_1^0=100$, $e_2^0=100$, $k_1=1$, $k_2=1$, $k_3=1$, 
$k_{-3}=1$, $k_4=2$ and $k_{-1}=1$.}
\label{fig:18}
\end{figure}

We conclude our analysis by noting that the conditions $\max \{\varepsilon_2,\mu_1,\mu_2\} \ll 1$ 
do not provide a \textit{universal} set of qualifiers to ensure that the phase-plane trajectory
approximately adheres to $\boldsymbol{x}^*$ after a brief fast transient. To establish criteria 
that determines a region in parameter space within which our analysis is valid, we first remark 
that an absolutely necessary condition for the validity of our timescale analysis is 
$V_2 \gg k_2c_1^{\max}$. Second, if $V_2 \gg k_2c_1^{\max}$ holds, then 
$0<t_{s_2} < t_{s_2}^{\chi}$ since
\begin{equation}
    t_{s_2}^{\chi} = t_{s_2}(1+\kappa_2)\left[1+\theta + \mathcal{O}(\theta^2)\right], \quad \theta \equiv \cfrac{k_2c_1^{\max}}{V_2}.
\end{equation}
Consequently, we take
\begin{subequations}
\begin{align}
    0 < &\min \{t_{s_2}^{\chi}/t_{s_1},t_{c_2}/t_{s_1}\},\\
    &\max \{t_{s_2}^{\chi}/t_{s_1},t_{c_2}/t_{s_1}\} \ll 1,
\end{align}
\end{subequations}
as our qualifying set of conditions that must hold in order for the trajectory to closely 
follow $\boldsymbol{x}^*$. This implies that the natural scaling to employ is given 
by (\ref{eq:sD3})--(\ref{eq:sD4}), and gives a universal set of parameters from which to 
analyze the phase-plane dynamics. For example, if $t_{s_2}^{\chi}\ll t_{s_1}$ but $\mu_2 \sim 1$, 
then we do not expect the trajectory to closely follow $\boldsymbol{x}^*$. However, we see 
from the scaled equations that $s_2$ should deplete in a QSS over the $t_{s_1}$ timescale 
as long as $t_{s_2}^{\chi}\ll t_{s_1}$. Thus, the trajectory $s_2$ ``sticks" to the 
$s_2$-nullcline, but lags behind $\boldsymbol{x}^*$ since $\mu_2 \sim 1$ 
(see {\sc Figure}~\ref{fig:17}). On the other hand, when the phase-plane trajectory does closely follow $\boldsymbol{x}^*$, the scaling given by (\ref{eq:sD3})--(\ref{eq:sD4}) tells us the component that contributes most to the error in our approximation (see {\sc Figure}~\ref{fig:18}).

\section{Discussion}
Enzyme catalyzed reactions typically exhibit multiple dynamical regimes; each regime marks 
a domain over which certain kinetic behavior and approximate rate laws can be assumed to be 
valid. The approximate rate laws are derived assuming timescale separations.  The primary 
contribution of this paper is to categorize specific types of timescales, particularly with 
regard to matched asymptotics in enzyme catalyzed reactions. In short, we have shown that 
in each kinetic regime of a reaction there really exist two distinct timescales that must 
be considered: characteristic and matching. Characteristic timescales arise naturally when 
the initial fast transient of a reaction can be approximated with a linear equation. This 
happens often in enzyme catalyzed models, since the differential equation governing the 
fast variable becomes linear when the slow variable is held constant. As such, the 
characteristic timescale should be utilized in scaling analysis, since it determines the 
relevant length scale of its corresponding regime. However, its limitation resides 
in the fact that it does not provide a good approximation to the time it takes a reaction 
to reach QSS. The matching timescale provides a reliable estimate to reach QSS, and determines 
the temporal boundary of the corresponding regime.

In this work, the fast and slow timescales of the single-enzyme, single-substrate MM reaction 
mechanism (\ref{eq:react1}) have been revisited. Under the RSA, the established fast timescale, 
$t_{c_1}$, of the MM reaction mechanism is a characteristic timescale: it provides the temporal 
order of magnitude needed for the concentration of complex to accumulate to approximately 
$63 \%$ of its threshold value. This is \textit{the} appropriate timescale to utilize in the 
scaling analysis. However, since $t_{c_1}$ does not provide a good estimate of when the 
complex concentration reaches its maximum value, it fails to define an appropriate matching 
timescale. The matching timescale delimits the approximate time point in the course of the 
reaction when the transition from initial fast transient to steady-state kinetics occurs. 
By utilizing Tikhonov/Fenichel theory, we have shown that the appropriate matching timescale 
for the MM reaction mechanism is $t_{c_1}^{\ast}$:
\begin{equation}
\nonumber t_{c_1}^{\ast} = -t_{c_1}\ln \cfrac{t_{c_1}}{t_{s_1}}.
\end{equation}

In this paper, we consider the auxiliary enzyme reaction 
mechanism~(\ref{eq:react2})--(\ref{eq:react4}) as a multiple timescale case study.  This
reaction was initially analyzed with the assumption that the auxiliary enzyme concentration is high, and that the primary reaction obeys the RSA. We demonstrated that when the secondary 
reaction has sufficient speed, the overall kinetics and reaction mechanism is determined by 
the ratios of four timescales: $t_{c_1}$, $t_{s_2}$, $t_{c_2}$ and $t_{s_1}$. Six different 
orderings of these timescales were considered: (i) $t_{c_1}\ll t_{s_2}\ll t_{c_2}\ll t_{s_1}$,
(ii) $t_{c_1}\ll t_{c_2}\ll t_{s_2}\ll t_{s_1}$, (iii) $\{t_{c_2},t_{s_2}\} \ll t_{c_1}\ll t_{s_1}$,
(iv) $t_{c_2}\ll t_{c_1} \ll t_{s_2}\ll t_{s_1}$, (v) $t_{s_2}\ll t_{c_1} \ll t_{c_2}\ll t_{s_1}$,
and (vi) $t_{c_1}\ll t_{s_2} = t_{c_2}\ll t_{s_1}$. The lag time, which is roughly the time 
it takes for the rate of product generation to reach its maximum value, was calculated for each specific ordering. As we have shown, the lag time corresponds to a specific matching timescale; specifically, we have demonstrated that the lag time is synonymous with the matching timescale that corresponds to the slow variable when the auxiliary reaction is composed of super-fast, fast, slow and super-slow variables.  
 
The estimation of timescales is perhaps the most challenging component chemical kinetics. 
The subtle difference between characteristic and matching timescales is often neglected 
in applications of GSPT. This work provides a useful case study in the interpretation of 
timescales in enzyme-catalyzed reactions, and the approaches used should be readily 
applicable to a wide range of singular perturbation problems in mathematical biology.

On a final note, we wish to emphasize that we carried out this analysis by restricting 
the parameters pertinent to the primary reaction to lie in a regime in which the RSA and 
QSSA are applicable. This is of course not necessary, and the total quasi-steady-state 
approximation could have been 
employed \citep{BORGHANS199643, SCHNELL2002137, Tzafriri2003,Bersani2012,Bersani2015}. 
The tQSSA is lumping method that it is generally considered to be valid over a much 
larger parameter range than the QSSA. It has been preivously applied to complex 
enzyme catalyzed reactions that exhibit both reversibility \citep{TZAFRIRI2004303} 
and competition \citep{Pedersena2006}. From a timescale perspective, the tQSSA has an 
advantage it reduces the total number of timescales in the system by lumping two chemical
species into one by defining the total substrate $s_T=c_1 + s_1$. The disadvantage of 
this approach is that the lumping of variables inevitably leads to a lower dimensionality 
system with a potentially different dynamical behavior. So far, the validity and applicability
of the tQSSA in the case of both the auxiliary reaction and coupled zymogen activation 
reactions \citep{EILERTSENBPC} remains open, and we certainly encourage exploration and 
research in this direction. 

\section*{Acknowledgements}
This work is partially supported by the University of Michigan Protein Folding Diseases 
Initiative, and Beilstein-Institut zur F\"{o}rderung der Chemischen Wissenschaften through 
its Beilstein Enzymology Symposia. We are  grateful to Antonio Baici (University of Zurich) 
for helpful discussions about this work during the 2017 Beilstein Enzymology Symposia 
(R\"{u}desheim, Germany. WS is a fellow of the Michigan IRACDA program (NIH/NIGMS 
grant: K12 GM111725).

\newpage








\end{document}